# Digital transformation: A systematic review and bibliometric analysis from the corporate finance perspective

Ping Zhang[1], Yiru Wang[2]

[1]*School of Finance, Capital University of Economics and Business, Beijing, China;*

[2]*School of Finance, Nankai University, Tianjin, China;*

Ping Zhang
School of Finance
Capital University of Economics and Business , China
**E-mail:** zhangping@cueb.edu.cn

Yiru Wang
School of Finance
Nankai University, China
E-mail: wangyiru_nku@163.com

# Digital transformation: A systematic review and bibliometric analysis from the corporate finance perspective

**Abstract:** Digital transformation significantly impacts firm investment, financing, and value enhancement. A systematic investigation from the corporate finance perspective has not yet been formed. This paper combines bibliometric and content analysis methods to systematically review the evolutionary trend, status quo, hotspots and overall structure of research in digital transformation from 2011 to 2024. The study reveals an emerging and rapidly growing focus on digital transformation research, particularly in developed countries. We categorize the literature into three areas according to bibliometric clustering: the measurements (qualitative and quantitative), impact factors (internal and external), and the economic consequences (investment, financing, and firm value). These areas are divided into ten sub-branches, with a detailed literature review. We also review the existing theories related to digital transformation, identify the current gaps in these papers, and provide directions for future research on each sub-branches.

# 1. Introduction

The transformation of production methods driven by big data and cloud computing, the innovation of transaction models led by digital finance and platform transactions, and the networked circulation of production factors are manifestations of a digital revolution. This revolution is prompting countries to accelerate their digital transformation strategies. According to the Global Digital Economy Development Index Report, the digital economy index rose from 45.33 in 2013 to 57.01 in 2021, reflecting a growth rate of 26%. These digital advancements have placed unprecedented competitive pressure on firms, demanding rapid adaptation to shifting market conditions. In particular, the widespread adoption of digital payments has heightened the need for enhanced innovation and resilience (Pan et al., 2022; Zhang et al., 2024). Alibaba stands out as a leading example of corporate digital transformation. By building a data platform, it has broken down information silos and improved its innovation capabilities. Alibaba's ability to swiftly leverage its digital resources to address external uncertainties, such as the pandemic, highlights its digital resilience. It is evident that digital transformation has revolutionized market expansion strategies and become essential for corporate survival and sustainable growth.

Digital transformation, which emphasizes data as the new production resource, digital space as the new growth domain, and data assets as the latest source of value, advances high-quality corporate development. Digital transformation reshapes industries, and scholars have increasingly focused on understanding its broader implications. A review of digital transformation research shows that academic studies in this field have increased annually. Existing studies remain relatively fragmented and have not reached a consensus on digital transformation's impact factors and economic consequences (Nadkarni & Prügl, 2021).

Specifically, the limitations of existing studies are mainly reflected in the following three aspects. There remains a debate over how to measure corporate digital transformation. Some studies argue that digital investment demonstrates the implementation and priority of a firm's strategy and should be measured from the perspective of digital R&D investment (Xu et al., 2023). However, some studies reveal that investments can be manipulated by firms, leading to earnings management behaviors such as "R&D manipulation" or "window dressing" of financial statements (Li et al., 2023). Vial (2019) emphasizes the application and practice of digital transformation, advocating for measuring digital transformation from digital patents. Recently, scholars have focused on text analysis's advantages in variable measurements, such as information richness, dynamic updates, and complex correlations. They tend to use text analysis to measure digital transformation. Therefore, accurately identifying digital transformation is a key issue (Luo, 2022).

In 2021, McKinsey surveyed over 800 firms worldwide and found that while 70% had initiated digital transformation, 71% remained stuck in the pilot stage. Addressing the incentives for digital transformation is essential to overcoming this dilemma. Existing literature on the drivers of digital transformation typically focuses on either internal governance effects or external monitoring effects. For example, Suppliers and customers are important enterprise stakeholders. Rising expectations for digital experiences and changing customer demands prompt firms to accelerate strategic adjustments, inevitably increasing pressure to speed up digital transformation (Geng et al., 2024). CEOs are the primary decision-makers. Firms led by CEOs in STEM exhibit more significant innovation and generate more digital patents (Kong et al., 2023). While previous studies have well documented these external and internal factors, few have systematically integrated them to analyze their heterogeneity or propose future research directions.

Many studies have highlighted that digital transformation has a strong potential in corporate investment and financing (Niu et al., 2023; Zhou & Li, 2023), productivity (Nucci et al., 2023; Wang et al., 2023), and value creation (Bresciani et al., 2021; Hadjielias et al., 2021). However, digital transformation may also negatively affect organizational structure and short-term performance. Zhong & Ren (2023) find that digital transformation firms often do not compensate for the costs they incur, thus worsening their business performance. Feliciano-Cestero et al. (2023) provide a systematic literature review to show digital transformation can positively and negatively impact firm internationalization at the individual, firm, and macro levels. Therefore, we believe that scholars have shown conflicting empirical evidence on the economic consequences of digital transformation due to the following two factors. Firstly, they primarily considered samples of firms from developing economies (such as China) rather than developed economies. Secondly, few studies consider the impact of digital transformation from multiple perspectives, but rather from a single perspective. From the corporate investment perspective, Li et al. (2023) find that digital transformation improves capital allocation efficiency by reducing agency costs and enhancing operational capabilities (Zhou & Ge, 2023). Guo et al. (2023), focusing on corporate financing, show digital transformation significantly alleviates financing constraints for small and medium-sized enterprises by improving the quality of information disclosure, thereby supporting their long-term stable development (Jiang et al., 2024).

To address this gap, in this study, we aim to provide a state-of-the-art review of the existing literature to identify the central issues around digital transformation. We propose the following research questions:

*RQ1. What are the theories, theoretical constructs, methodologies, and contexts of interest examined by prior researchers in digital transformation and its effects on firms' economic consequences?*

*RQ2. What is the measurement of previous researchers for digital transformation? What factors motivate or hinder a firm's digital transformation?*

*RQ3. How does digital transformation impact firms' investment, financing or value creation?*

This study contributes to the literature in three ways. First, it provides a systematic review within the corporate finance framework, expanding our understanding of digital transformation. Existing reviews on digital transformation mainly focused on a management perspective (Nadkarni & Prügl, 2021; Isensee et al., 2020; Parra-Sánchez & Talero-Sarmiento, 2024) and failed to summarize digital transformation within the corporate finance framework (see Figure 1). This paper offers a comprehensive synthesis from three aspects: the measurement of digital transformation, internal and external factors influencing digital transformation, and the economic consequences of digital transformation, including its impact on corporate investment, financing, and firm value. Through this analysis, we provide a more detailed understanding of the role of digital transformation in corporate finance. Second, this paper explores the multiple impacts of digital transformation strategies on firm behavior. By organizing research based on the aforementioned three perspectives, we further refine the analysis into specific sub-branches: internal and external factors affecting digital transformation, digital transformation and corporate investment (including investment strategies and investment efficiency), digital transformation and corporate financing (including financing constraints, channels, and capital structure), and digital transformation and corporate value (including total factor productivity and market valuation). Additionally, we link these sub-branches to relevant economic theories, offering a clear framework and theoretical support for future research. Third, this paper highlights future research opportunities in each branch of digital transformation. In particular, researchers have largely overlooked areas such as integrating digital transformation and green development, data ownership, and intellectual property protection. A critical future research direction lies in incorporating digital resources as capital into the firm's production function and analyzing how these resources contribute to sustainable development through interactions across multiple economies.

The remainder of this paper is structured as follows: Section 2 describes the methodology of the literature search and analysis techniques applied in this study. Section 3 conducts descriptive statistics on digital transformation literature, including the number of publications, institutions, countries, etc. In Section 4, the authors conduct a scientific mapping of the knowledge domain. This endeavor consists of co-occurrence analysis, co-word analysis, co-citation analysis (cluster analysis), and content analysis of each sub-branches. Section 5 presents an in-depth discussion, evaluation of the key debates and research gaps in the field, and a summary of future trends and research opportunities in the field.

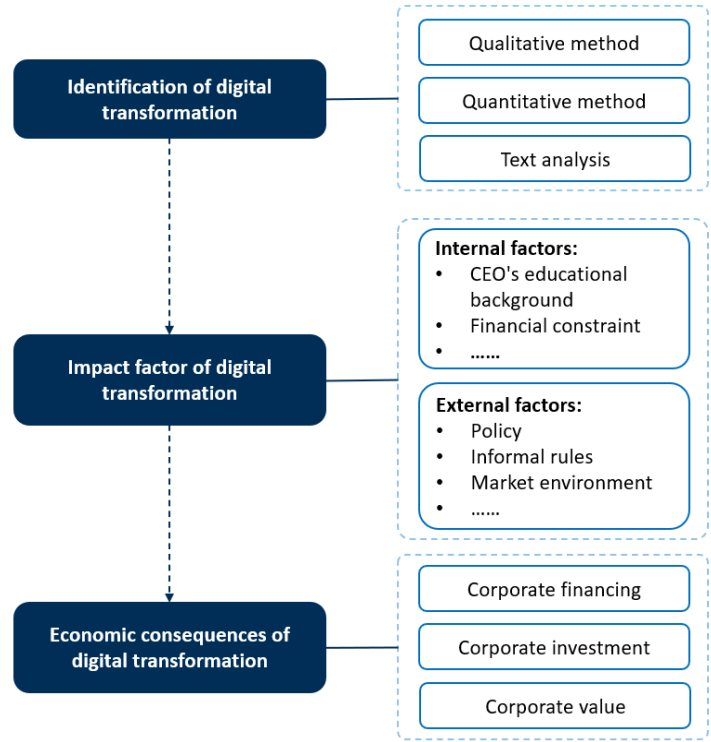

Fig. 1 The framework of this paper.

## 2. Data sources and research methodology

2.1 Data sources

The literature sample used for this study is sourced from Web of Science. To ensure the quality of these papers, we select the SSCI and SCI-E subsets as data sources. To answer the research questions, we collected 1399 pieces from 2011 to 2024 using the following keywords: "Digital transformation" AND "Investment," "Digital transformation" AND "Financing," and "Digital transformation" AND "Firm value." Specifically, there are 667 papers on "Digital transformation" AND "Investment," 447 papers on "Digital transformation" AND "Financing," and 443 papers on "Digital transformation" AND "Firm value."

2.2 Research methodology

We employ bibliometric and content analysis methods to summarize the characteristics of digital transformation research. Bibliometrics, combining mathematics, statistics, and bibliography, provides a quantitative overview of knowledge (Ninkov et al., 2022), offering an objective understanding of the knowledge structure and research development in a field. However, it lacks depth in analyzing specific content. Content analysis, a common review method, offers insights into more particular research by organizing and summarizing literature (Vaismoradi et al., 2013). Still, it has limitations such as subjectivity and limited literature.

Therefore, we adopt a comprehensive quantitative and qualitative research methodology based on data mining and metrological analysis of digital transformation literature in terms of "Institution," "Country," cited literature, citations, etc., we distill the knowledge base of digital transformation research field through content analysis to grasp the development pulse, hotspot changes, and future trends in the digital transformation field, and to provide a basis for related researches. Fig. 2 represents the detailed flow chart of the review process employed in this study.

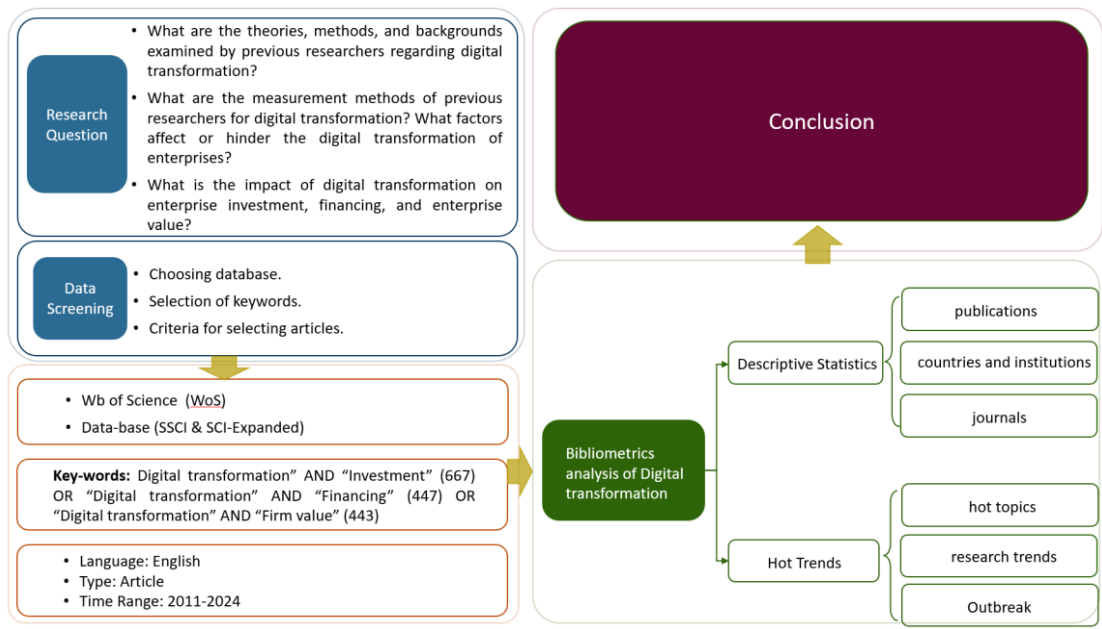

Fig 2 Flow chart of review on digital transformation.

## 3. Digital transformation research characteristics

3.1 Trend in publications on Digital Transformation

Fig. 3 shows the distribution of publications over time, indicating a steady rise in digital transformation research. We identified three stages based on publication trends. The period from 2010 to 2015 marks the emergence of this field, with fewer than 20 annual publications reflecting limited attention. From 2015 to 2020, the field developed significantly, with publications reaching around 100 by 2020. This increase is likely linked to the OECD's 'Digital Economy Outlook' reports released in 2015 and 2017, which outlined trends and policies in the digital economy. 2016, the European Digital Industrial Strategy was launched to integrate the digital strategies of EU member states. Since 2020, the number of publications has surged, reaching 505 by 2023, making it a hot topic among scholars.

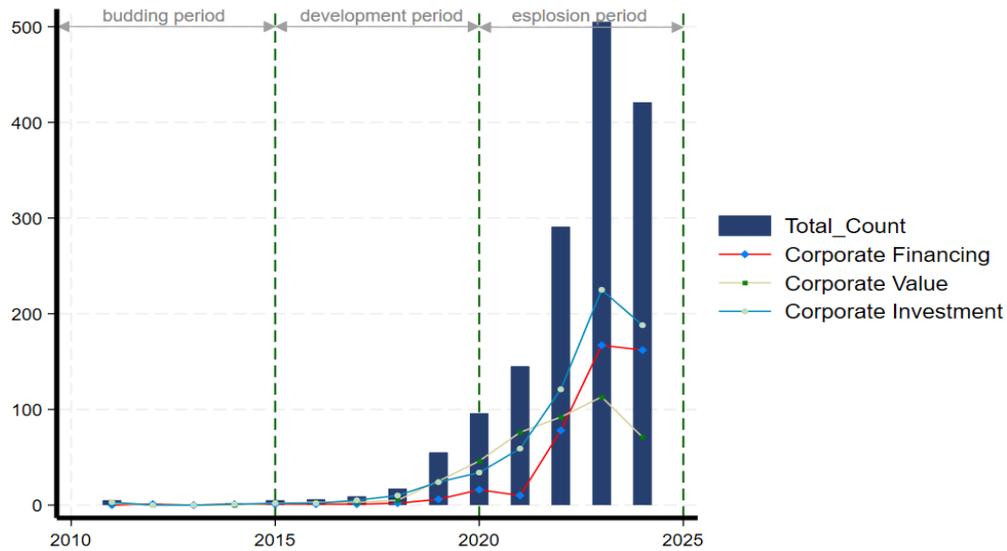

Fig. 3 Annual change in the number of publications frequency in digital transformation.

3.2 Analysis of research strengths and collaborations on digital transformation

3.2.1 Analysis of inter-country cooperation relations

This paper conducts a social collaboration network of countries in digital transformation research, using k=25 to generate a collaboration network visualization (see Fig. 4). It can be seen that the number of Chinese paper outputs is in the first place, which is consistent with its "Digital China" initiative. As of the search date, China has published 712 papers on "digital transformation" in corporate finance, reflecting its significant contribution. The U.S. follows with 139 papers, the U.K. with 112, Italy with 108, and Germany with 79. Notably, only two developing countries—China and India—are in the top ten, indicating that developed countries dominate digital transformation research.

Centrality measures the shortest paths passing through a node, reflecting its importance in the network's connectivity. Fig. 4 shows that Germany and France have numerous connections with other countries. As indicated in Table 1, Germany has the highest centrality (0.45), followed by France (0.44) and Spain (0.41), reflecting a high level of openness in research collaboration between countries and their institutions. This also highlights these countries' strong influence and international discourse power in digital transformation. While China performs strongly in terms of publication volume, the centrality of its research remains low, pointing to limited collaboration in digital transformation studies. To further develop this field, Chinese scholars should enhance cooperation and communication with domestic and international research institutions.

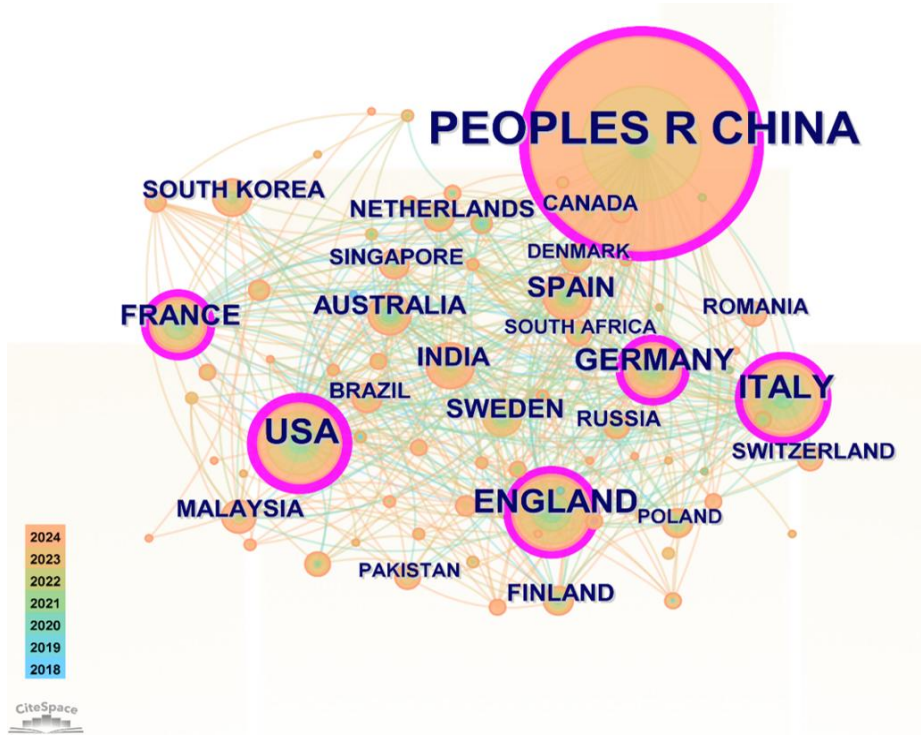

Fig. 4 Analysis of inter-country cooperation relations

Table 1 Analysis of inter-country cooperation relations

| No. | Freq | Burst | Centrality | Label |
| --- | --- | --- | --- | --- |
| 1 | 712 | 0 | 0.11 | PEOPLES R CHINA |
| 2 | 139 | 9.24 | 0 | USA |
| 3 | 112 | 3.15 | 0.11 | ENGLAND |
| 4 | 108 | 5.38 | 0.21 | ITALY |
| 5 | 79 | 6.32 | 0.45 | GERMANY |
| 6 | 61 | 0 | 0.44 | FRANCE |
| 7 | 51 | 3.47 | 0.41 | SPAIN |
| 8 | 43 | 6.41 | 0.16 | AUSTRALIA |
| 9 | 41 | 0 | 0.11 | INDIA |
| 10 | 40 | 4.6 | 0.36 | SWEDEN |

3.2.2 Analysis of inter-institution partnerships

Fig. 5 presents a social collaboration network of institutions in digital transformation research. Leading Chinese institutions like Renmin University of China and Shanghai University of Finance and Economics have prominent nodes. International universities like the University of London and the University of California also display high centrality, reflecting their strong collaboration with other institutions in this field. Unlike developed institutions, Chinese universities' networks are relatively broad but lack close cooperation with international research institutions, suggesting potential barriers to collaboration between developed and developing countries in digital transformation.

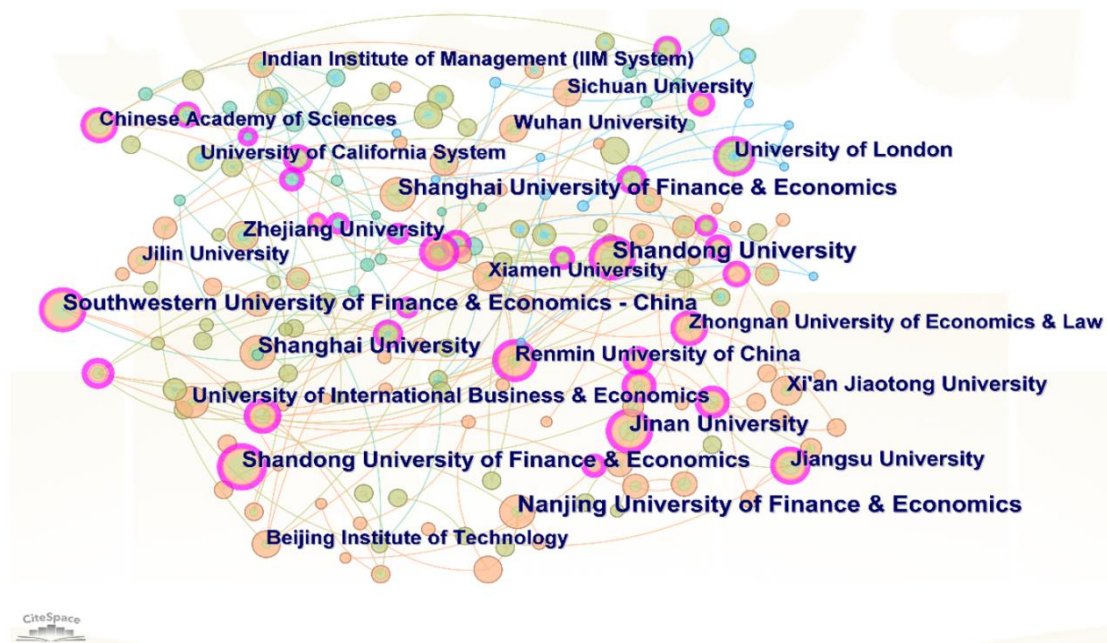

Fig. 5 Analysis of inter-institution partnerships

3.3 Journal of publication

Fig. 6 and Table 2 present a visualization of journals' co-citation network and a statistical analysis table based on co-citation frequency and centrality. The journal with the highest co-citation frequency was the Journal of Business Research, with 717, followed by the Technology Forecast and Social and Journal Clean Production. The nodes highlighted in the center color of Figure 6 have high centrality. Mis Quarterly, Journal Clean Production, and Energy Economics have high co-citation frequency and centrality, indicating that these journals have made significant theoretical contributions to the development of digital transformation research.

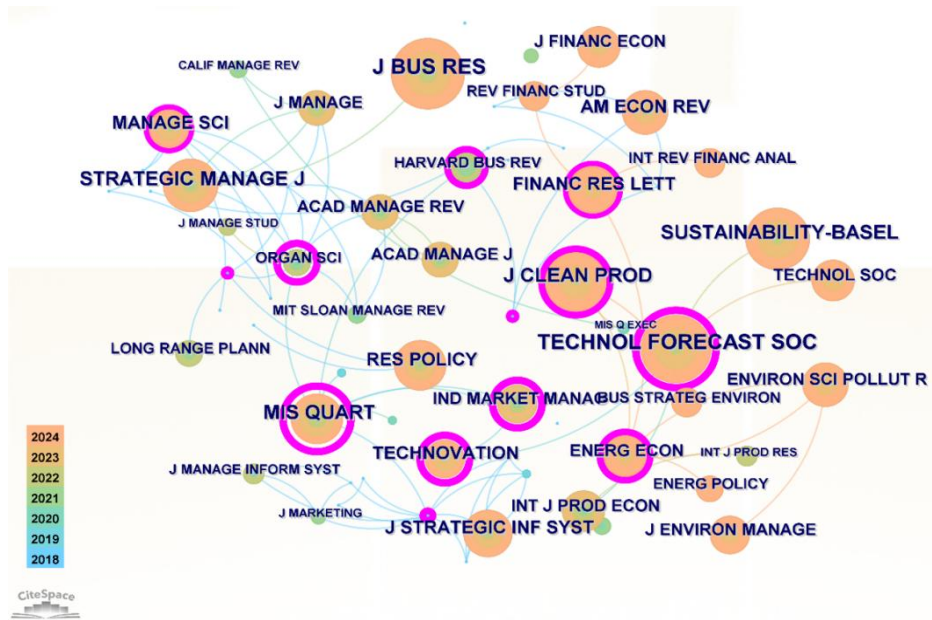

Fig. 6 Journal of publication.

Table 2 Journal of publication.

| No. | Freq | Centrality | Journal |
| --- | --- | --- | --- |
| 1 | 717 | 1.1 | J BUS RES |
| 2 | 713 | 1.04 | TECHNOL FORECAST SOC |
| 3 | 512 | 0 | SUSTAINABILITY-BASEL |
| 4 | 463 | 0.63 | J CLEAN PROD |
| 5 | 439 | 1.17 | STRATEGIC MANAGE J |
| 6 | 355 | 0 | RES POLICY |
| 7 | 309 | 0.69 | MIS QUART |
| 8 | 281 | 0 | AM ECON REV |
| 9 | 176 | 0.29 | FINANC RES LETT |
| 10 | 176 | 0 | IND MARKET MANAG |
| 11 | 173 | 0 | INT J PROD ECON |
| 12 | 167 | 0 | J STRATEGIC INF SYST |
| 13 | 125 | 0.55 | ENERG ECON |
| 14 | 124 | 0 | ENVIRON SCI POLLUT R |
| 15 | 117 | 0.15 | J FINANC ECON |

# 4. Digital transformation research hotspots and stages frontier analysis

4.1 Analysis of research hotspots

　　Keywords provide a concise content summary, clearly reflecting the paper's theme and core ideas. In the literature, the frequency of keyword occurrence indicates the research hotspots and trends in the field; a higher frequency suggests more prominent research interests. We extracted keywords with a co-occurrence frequency of over 50

times and listed them in Table 3. It can be seen that digital transformation is the keyword with the highest frequency of occurrence, with a frequency of 511 times, accounting for one-third of the entire sample, followed by innovation, technology, etc. Financing, investment, and firm value are also included in the hot topic word set, which verifies the efficiency of the classification in this review. In Table 3, we find that the centrality of hot keywords is relatively high. Keywords such as technology, firm performance, and strategy are key nodes in the foundation of research and interdisciplinary research. Fig. 7 presents the co-occurrence frequency of keywords in a network, consistent with the findings in Table 3. Keywords such as "digital technology" and "corporate performance" are closely linked to other keywords. This indicates that these topics are currently hot issues in the research field.

Table 3 Analysis of keyword occurrence

| No. | Freq | Centrality | Year | Keywords |
| --- | --- | --- | --- | --- |
| 1 | 511 | 0.26 | 2018 | Digital transformation |
| 2 | 300 | 0.03 | 2018 | Innovation |
| 3 | 204 | 0.97 | 2018 | Technology |
| 4 | 196 | 0.53 | 2018 | Performance |
| 5 | 189 | 0.00 | 2019 | Impact |
| 6 | 175 | 0.00 | 2019 | Management |
| 7 | 148 | 0.22 | 2019 | Transformation |
| 8 | 108 | 0.37 | 2018 | Strategy |
| 9 | 105 | 0.20 | 2020 | Dynamic capability |
| 10 | 97 | 0.00 | 2019 | Information technology |
| 11 | 96 | 0.29 | 2019 | Capability |
| 12 | 95 | 0.72 | 2019 | Firm performance |
| 13 | 91 | 0.25 | 2019 | Big data |
| 14 | 88 | 0.11 | 2021 | Investment |
| 15 | 81 | 0.03 | 2022 | Growth |
| 16 | 73 | 0.17 | 2022 | Digital finance |
| 17 | 69 | 0.25 | 2018 | Information |
| 18 | 67 | 0.00 | 2018 | Firms |
| 19 | 67 | 0.06 | 2022 | Value creation |
| 20 | 67 | 0.00 | 2021 | Model |
| 21 | 63 | 0.10 | 2022 | Financing constraints |

Fig. 7 Analysis of keyword occurrence.

Fig. 8 shows the keyword clustering map, with each color module corresponding to a clustering theme of digital transformation-related literature. Keyword clustering analysis reveals that research hotspots can be broadly categorized into three types: ① the concept and measurement of digital transformation (Cluster 3: dynamic capabilities, Cluster 5: quantitative measurement, Cluster 7: advanced technologies), ② factors influencing corporate digital transformation (Cluster 2: network infrastructure, Cluster 4: digital finance, Cluster 8: equity capital), and ③ economic consequences of digital transformation (Cluster 6: value creation, Cluster 9: business value, Cluster 10: carbon emission intensity).

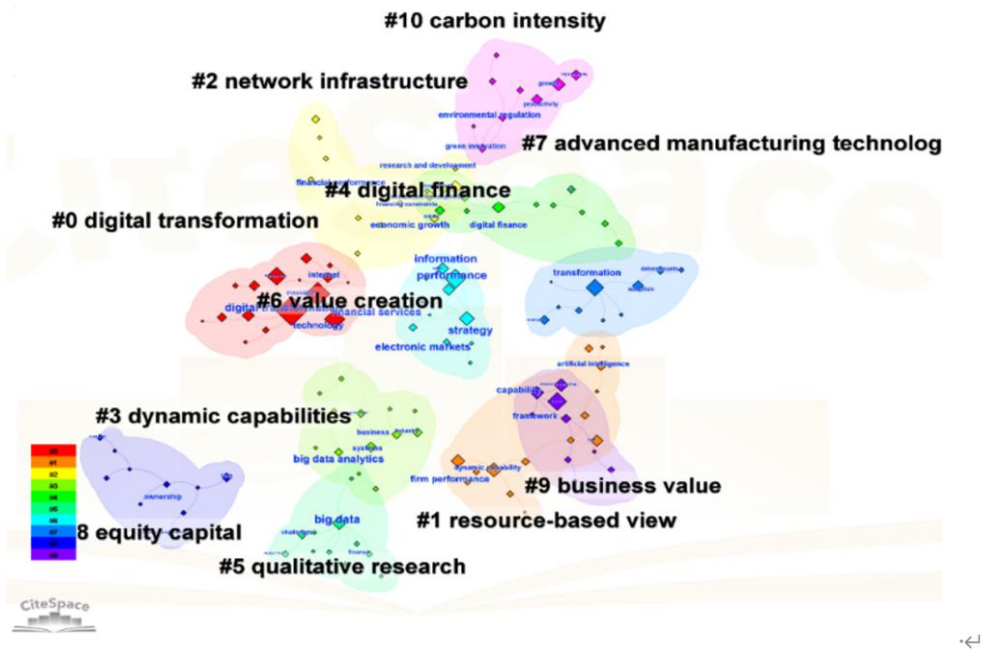

Fig. 8 Keyword clustering map.

4.1.1 Measurement of digital transformation (Cluster 3, Cluster 5, Cluster 7)

Measuring corporate digital transformation poses a challenge due to the lack of standardized criteria for digital transformation. Specifically, there are qualitative or quantitative methods in the existing literature to measure digital transformation. In terms of qualitative measurement, Dou et al. (2023) introduce a dummy variable to indicate whether a firm undergoes digital transformation in a given year. Furthermore, to reflect the "intensity" of digital transformation in quantitative measurements, researchers have calculated the proportion of a company's software or hardware investments related to digital technologies to total assets (Müller et al., 2018; Chun et al., 2008). They have also measured the use of robots within companies based on survey data (Acemoglu & Restrepo, 2020) and analyzed the use of General purpose technologies (GPTs) (Brynjolfsson et al., 2021). Text analysis has gradually become the mainstream method in measuring digital transformation in recent years. Build a keyword dictionary containing various digital technologies, and then construct enterprise digital transformation indicators based on the frequency or proportion of these keywords appearing in the "Management Discussion and Analysis" section of the firm's annual report.

The underlying assumption of this method is that mentioning keywords related to a specific digital technology in the text indicates that a firm has undergone digital transformation. There is much literature using this method. Xue et al. (2022) search the websites of the Central People's Government and the Ministry of Industry and Information Technology to obtain 30 important national-level digital economy-related policy documents for extracting keywords related to corporate digitization. After the Python word separation process and manual recognition, 76 terms related to enterprise digitization are finally screened and retained, constituting the dictionary of digitization terms (Shang et al., 2023). They use the natural logarithm of the total word frequency of digital keywords in the annual report, which provides novel evidence on a previously under-explored digital transformation measurement. The Chinese Corporate Digital Transformation Index in the CSMAR database is constructed based on information from annual reports, fundraising announcements, and qualification certifications, which can largely reflect the operating characteristics and development direction of the firm, and help investors grasp the essence of the company's operational strategies, competitive advantages and industry trends. This index is calculated by the weighted sum of the six primary indicators. These primary indicators include strategy-driven (Strategy), technology-enabling (Technology), organization-enabling (Organization), environment-enabling (Environment), digital achievement (Achievement), and digital applications (Application). These primary indicators comprise 31 secondary indicators, establishing the evaluation framework of corporate digital transformation.

It is worth noting that the text analysis may be more vulnerably affected by "linguistic manipulation" and "window dressing." On the one hand, digital technology keywords that can be included in the dictionary are not comprehensive enough, resulting in some digital transformations not being counted (firms apply digital technology but are not recognized by dictionary methods). On the other hand, a part of the text content is mistakenly judged as digital transformation (keywords are mentioned in the text, but digital technology is not used).

4.1.2 Factors influencing digital transformation (Cluster 2, Cluster 4, Cluster 8)

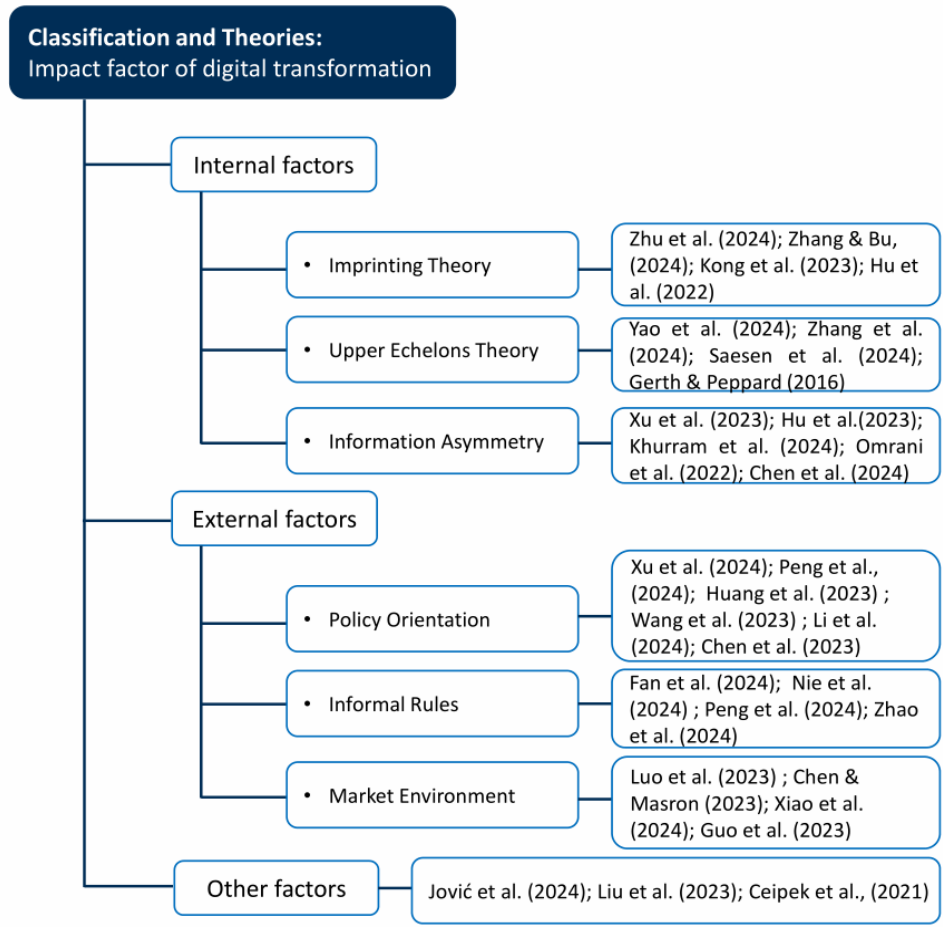

Fig. 9 Impact factors of digital transformation.

(1) **Internal factors**

Based on imprinting theory, Zhu et al. (2024) propose that IT top managers significantly facilitate the digital transformation of small and midsize enterprises (Zhang & Bu, 2024). Similarly, Kong et al. (2023) argue that firms led by CEOs with STEM (Science, Technology, Engineering, and Mathematics) backgrounds perform better in digital transformation. A CEO with a STEM degree is more adept at recognizing the value of R&D investments and the efficiency of innovative outputs, thereby enhancing the transition from technological innovation to achievements. Hu et al. (2022) find executives' overseas study and work experience enhanced enterprise digital transformation significantly, thus improving enterprise growth.

Based on upper echelons theory, Yao et al. (2024) suggest that CFOs' narcissistic tendencies promote a risk-taking spirit, more substantial decision-making power, and higher expectations for decision-making results, thus promoting corporate digital transformation. Zhang et al. (2024) reveal that concentrated equity weakens management's power stability, constrains their residual control, suppresses power, and undermines the effect of management's power on driving corporate digital transformation, based on the approach-inhibition theory of power. Adopting a digital orientation (DO) is a prerequisite for firms intending to lay the basis for digital strategic initiatives (Saesen et al., 2024). CEO overconfidence positively relates to a firm's digital orientation, particularly in industries with low market turbulence and technological dynamism. Gerth & Peppard (2016) review the topic of Chief Information Officers (CIOs) and digital transformation, elucidating how CIOs can avoid digital transformation failures. These actions include (1) understanding the CEO's vision for IT, (2) Understanding the ambiguity of the role of Chief Information Officer, (3) Fulfilling service and solution commitments, (4) Developing a relationship strategy, (5) Actively defining IT success; (6) Manage the process of change, etc. Jović et al. (2024) conduct case studies and interviews, finding that information security, investment in emerging technologies, leadership motivation, and expertise are crucial for digital transformation.

Based on information asymmetry, Xu et al. (2023) propose that the more significant the financing pressure on a company is, the worse the digital transformation and financing constraints exacerbate this negative effect. Building on the factors influencing corporate financing, Hu et al. (2023) suggest that corporate maturity mismatch hinders enterprise digital transformation significantly by exacerbating financial distress and distorting investment behavior. Additionally, investment characteristics are also related to digital transformation. Outward foreign investment performance significantly impacts firms' digital transformation (Khurram et al., 2024). Chen et al. (2024) revealed that corporate financialization has a negative effect on digital transformation and that the relationship is more pronounced in low-productivity firms, traditional industries, and high bankruptcy-risk groups. Omrani et al. (2022) aim to identify and analyze factors determining the adoption of digital technologies in SMEs. Drawing on the technology–organization–environment framework, the study highlights that the technology context (IT infrastructure and digital tools) and innovation are the main drivers that act as stepping stones in digital technology adoption. Corporate regulation, available skills, and financial resources (as organizational variables) also play a significant role in the adoption decision. The influence of the environmental context is marginal.

(2) **External factors**

The policy orientation plays a substantial role in guiding and supporting the digital transformation. Xu et al. (2024) show that the government's digital attention can promote enterprise digital transformation (Peng et al., 2024). In this process, the development of digital infrastructure and government investment serve as the mechanism channels. From government attention to behavior, Huang et al. (2023) find green finance policy is a stumbling block rather than a catalyst that hinders the digital transformation's enthusiasm of high pollution enterprises, and it plays an inhibitory role by increasing financing costs and financing constraints. Wang et al. (2023) divide government support into government subsidies and tax incentives and find that two forms of government support both positively promote enterprises' digital transformation. Li et al. (2024) find a non-linear relationship (inverted U-shaped) between government procurement and digital transformation. This effect is due to moderate government procurement promoting firm R&D investment and patents. However, excessive government procurement limits innovation performance and ultimately hinders digital transformation. Tax cuts are conducive to improving cash flow conditions and digital transformation. Chen et al. (2023) find business tax to value-added tax policy had significantly promoted digital transformation. Yang et al. (2023) show that expanding local government debt hinders corporate digital transformation. Such negative impact is more significant in non-state-owned enterprises and companies with low concentration of ownership and those in provinces with low fiscal self-sufficiency rates and low bank competition. Considering firms' preferences for risk strategies and the influence of leading firms (Zhu et al., 2023), there is heterogeneity in the incentive effects of government regulations on digital transformation. As a generalized environmental regulation policy at the city level, the low-carbon city pilot imposes carbon emission reduction constraints on all related subjects by formulating low-carbon development plans and bringing them into the government's major tasks. Zhao et al. (2023) find that low-carbon city pilots have promoted the digital transformation of manufacturing enterprises in pilot cities by strengthening fiscal expenditures on science and technology (S&T) and alleviating financing constraints. Wen & Deng (2023) propose that data is intellectual property. There is a significant positive relationship between intellectual property protection and the digital transformation.

From the perspective of informal institutions, various factors influence digital transformation. Public environmental attention is one of the representatives of informal ecological regulation. Fan et al. (2024) indicate that public environmental attention encourages companies to invest in research and development (R&D), attract skilled personnel, and commit to ecological responsibilities, thus promoting digital transformation. Conversely, Nie et al. (2024) find firms located in high social capital provinces, especially those with more advanced social networks, demonstrate a lower propensity for digital transformation. Additionally, venture capital, a key player in the capital market, is an essential external force influencing enterprise development. Peng et al. (2024) identify venture capital's three roles in digital transformation: financial assistance, governance empowerment, and capability support. Furthermore, corporate culture, another aspect of informal institutions, also impacts digital transformation. Zhao et al. (2024) reveal that a gambling culture hinders digital transformation by limiting talent development and innovation investment, which are crucial drivers of this process.

From the market environment perspective, Luo et al. (2023) highlight a better business environment has a positive relationship with corporate digital transformation. Attracting professional leadership from top managers, enhancing investment in digital technology, and increasing government subsidies related to digitalization provided by a better business environment are three possible channels that facilitate digital transformation. The macro environment uncertainty often affects enterprises' investment decisions. Chen and Masron (2023) find that economic policy uncertainty promotes corporate digital transformation by intensifying market competition. In contrast, Xiao et al. (2024) discover that air quality weakens digital transformation by increasing economic policy uncertainty, reducing human capital, and exacerbating the crowding-out effect on environmental investments (Wang et al., 2024).

More importantly, digital transformation is influenced by internal and external factors, as well as the incentive or pressure effects from associated firms. Guo et al. (2023) find major customers' digital transformation is positively related to their supplier's digital transformation, and the effect is more pronounced when customers are of greater importance to suppliers (Wang et al., 2024). Liu & Chen (2024) refined their analysis using a supply chain sample. They suggest that sharing auditors with customers can promote suppliers' digital transformation by strengthening the suppliers' supervision and alleviating their financing constraints. However, digital transformation requires relentless data capital investment (DCI), often constrained by a dilemma that imposes a stark trade-off between investment cost and returns. Neither the supplier nor the customer invests in data capital when the data capital investment is costly.

In contrast, both invest in data capital when the data capital investment benefits are relatively significant (Xin et al., 2024). Financing constraints are one of the barriers to digital transformation. Hua & Yu (2023) find that local competition can promote digital transformation by increasing digital investments and alleviating financing constraints. Bai et al. (2024) argue that competition in the banking branches can significantly boost digital transformation. This competition drives digital transformation by easing financing constraints, reducing operational risks, and increasing R&D investments.

(3) **Others**

Family firms have stronger needs for legitimacy and are sensitive about investing in uncertain projects (Liu et al., 2023). They are more likely than their non-family counterparts to disclose more symbolic cues about digital transformation while investing less substantively in digital transformation. That is, they emphasize digital transformation in their annual reports to a greater extent while making fewer digital investments. In addition, due to the unique characteristics of family management, such as non-economic goal orientation centered on the family, long-term tenure and emotional connections to existing assets. Family members in the senior management team can inhibit exploratory digital innovation (Ceipek et al., 2021).

4.1.3 Digital transformation economic consequences (Cluster 6, Cluster 9, Cluster 10)

4.1.3.1 Digital transformation and corporate financing

In this section, we analyze the 447 papers related to "digital transformation" and "corporate financing." Specifically, we categorize the literature into financing constraints, financing channels, and capital structure. Existing studies are based on resource-based, information asymmetry, and stakeholder theories to discuss digital transformation and financing constraints. For the financing channels, we classify the literature according to information asymmetry theory and risk-taking theory. Regarding digital transformation and capital structure, existing research analyzes these topics based on the Modigliani-Miller theorem and pecking order theory.

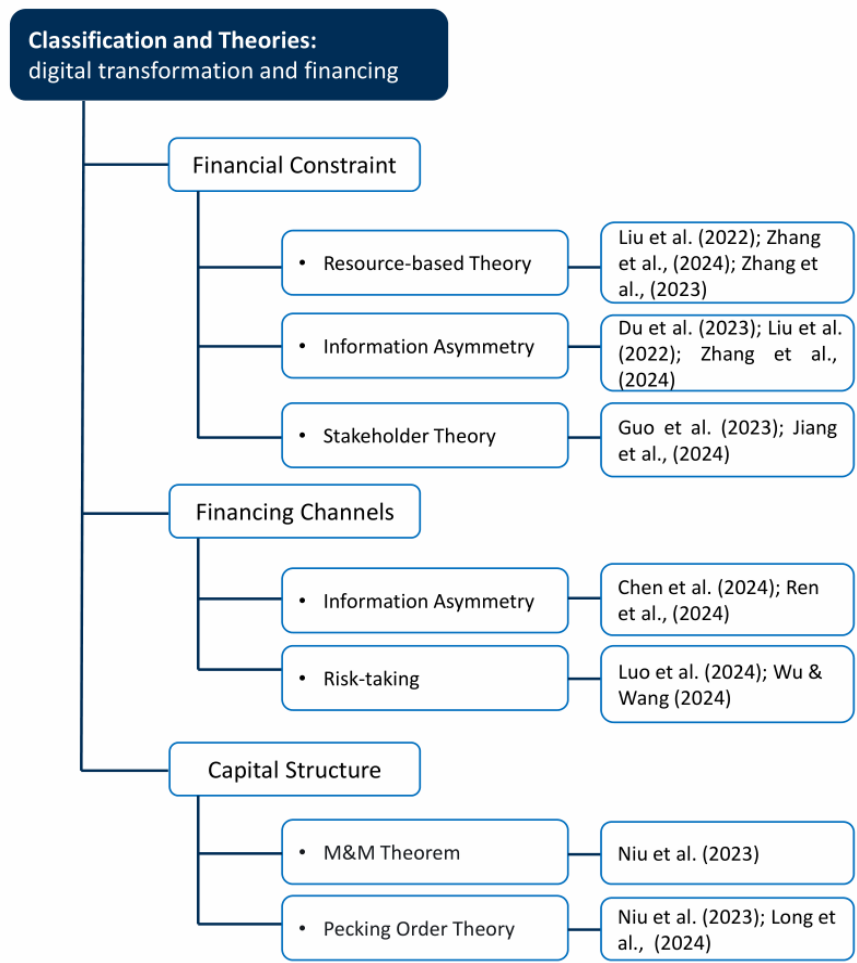

Fig. 10 Digital transformation and corporate financing.

**(1) Financing constraints**

Liu et al. (2022) argue that digital transformation promotes enterprise development by alleviating financing constraints, reducing business risks, and driving technological innovation (Zhang et al., 2024; Zhang et al., 2023). Furthermore, Du et al. (2023) find that digital network technology increases financing constraints initially but reduces them in the second lag period. Due to their scale and information disadvantage, small and medium-sized enterprises often face significant challenges in financing. Guo et al. (2023) discover SMEs improve the quality of external information disclosure through digital transformation, thereby alleviating financing constraints (Jiang et al., 2024).

**(2) Financing channel**

Diversified financing is one of the critical measures for firms to mitigate financial risks. Chen et al. (2024) argue digital transformation alleviates bank credit dependence mainly by easing firms' financing constraints and reducing information asymmetry. In this way, digital transformation can expand the financing channels for firms. And digital transformation significantly promotes risk-taking (Luo et al., 2024). The positive impact is more significant in non-heavily polluting, small-scale, and non-state-owned enterprises. Wu & Wang (2024) show that optimized corporate governance processes and increased investment in research and innovation act as positive intermediaries through which digitalization affects the level of corporate risk performance.

**(3) Capital structure**

M&M theorem posits that capital structure is irrelevant to firm value under strict assumptions. However, due to friction factors such as bankruptcy costs, agency costs, and information asymmetry, capital structure is linked to firm value. As a result, firms adjust their debt-to-equity ratios to reach an optimal level, minimizing financing costs and improving firm value. Digital transformation can improve the probability of leverage upward (downward) adjustment by issuing debts and dividends (issuing equities and repaying debts) (Niu et al., 2023).

4.1.3.2 Digital transformation and corporate investment

In this section, we analyze the 667 papers related to "digital transformation" and corporate investment." Specifically, we categorize the literature into investment efficiency and investment strategy. Regarding digital transformation and investment efficiency, we categorize the literature into inefficient investment, green investment efficiency, and labor investment efficiency. In terms of digital transformation and investment strategies, we analyze the literature based on investment horizons and investment types.

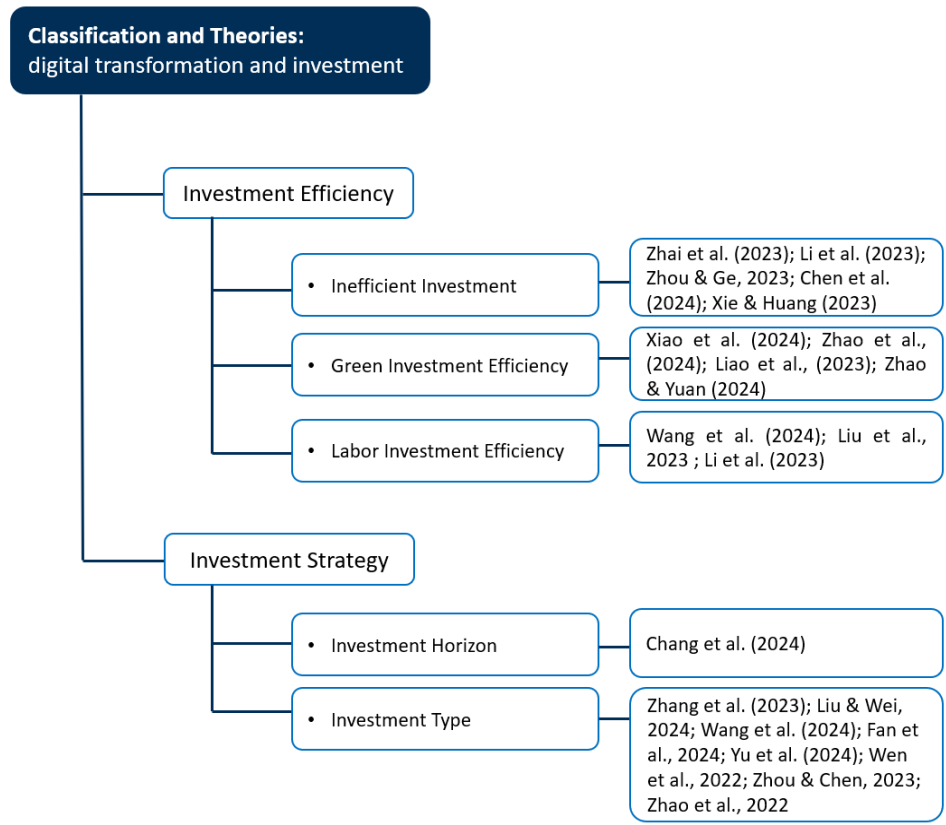

Fig. 11 Digital transformation and corporate investment.

(1) **Investment efficiency**

Inefficient investment refers to allocating excessive or insufficient resources and capital to projects beyond their economic benefits. It leads to misallocating resources, reduced operational efficiency, and declining firm performance and value (Piluso, 2024). Zhai et al. (2023) reveal that digital transformation effectively inhibits over-investment. Compared to firms without digital transformation, those undergoing or possessing a higher degree of digital transformation exhibit lower probabilities and degrees of over-investment. Li et al. (2023) find digital empowerment is a crucial measure to enhance capital allocation efficiency by reducing agency costs and enhancing operational capabilities (Zhou & Ge, 2023).

On the contrary, Chen et al. (2024) reveal that enhancing corporate digitalization substantially positively affects maturity mismatch because of the long-term and high-risk of digital technology. From the perspective of corporate taxation, Xie & Huang (2023) find that digital transformation significantly curbs tax avoidance by improving internal control quality, thereby enhancing investment efficiency. Xiao et al. (2024) show that digital transformation can promote the efficiency of green technological innovation under environmental regulations by reducing costs (Zhao et al., 2024; Liao et al., 2023).

From the human capital perspective, Wang et al. (2024) propose that digital transformation increases corporate labor investment efficiency by reducing agency problems and mitigating financing restrictions (Liu et al., 2023). Li et al. (2023), considering the relationship between digital transformation and income inequality, find that corporate digital transformation significantly reduces income inequality between managers and employees, but it also increases the income gap between high-skilled and low-skilled workers.

Considering the impact of associated enterprise, Zhao & Yuan (2024) find that digital transformation significantly promotes the herding effect in corporate green investment. Jiang et al. (2024) note that peer digital transformation substantially enhances the efficiency of corporate investment decisions. Moreover, narrowing the digital divide between industries and regions strengthens the positive impact of digital transformation on capacity utilization (Chen et al., 2024).

(2) **Investment strategy**

According to the investment horizon classification, Chang et al. (2024) demonstrate that digital transformation significantly reduces long-term firm investment. Industrial peers' digital transformation prohibits a focal firm's long-term investment.

According to the investment type classification, Zhang et al. (2023) find that digital transformation exerts a negative impact on corporate financialization by decreasing financial investments while increasing real investments. This is because digital transformation to redeploy a greater number of resources from financial assets to real investments.

More specifically, there is a significant inverted U-shaped relationship between digital transformation and defensive financial asset allocation and a significant positive relationship with profit-seeking financial asset allocation (Liu & Wei, 2024). Wang et al. (2024) indicate that digital transformation drives OFDI by reducing overseas operational costs and enhancing technological innovation capabilities (Fan et al., 2024; Peng et al., 2022). Yu et al. (2024) find that digital transformation can improve corporate innovation investment (Wen et al., 2022; Zhou & Chen, 2023; Zhao et al., 2022) and enhance technological innovation capabilities (Zhao et al., 2022; Du & Wang, 2024; Xu et al., 2024).

Addressing how digital transformation overcomes innovation challenges, Zhuo & Chen (2023) propose that digital transformation promotes corporate innovation by improving innovation quality and enhancing knowledge absorption capabilities. Zhang et al. (2023) find digital transformation increases the volatility of innovation investment. Environmental issues have steadily emerged, drawing increasing public attention. These include smog in cities, oil pollution, and more. Scholars have also increasingly focused on the impact of digital transformation on green investment and environmental performance. Jin et al. (2023) find that digital investment has a significant U-shaped relationship with corporate ecological performance, and technological innovation is an intermediary channel through which digital investment promotes enterprise environmental performance. Li et al. (2023) show that digital transformation mitigates agency problems by monitoring loan usage, overseeing green technology projects, and reducing managerial myopia, thereby easing financing constraints and promoting green innovation (Zhang et al., 2024; Xu et al., 2024; Liu et al., 2024; Feng et al., 2022; Dong et al., 2024; Liu et al., 2023; Tang et al., 2023; Du et al., 2023).

Chen et al. (2023) further classify green innovation quality and find that digital transformation primarily promotes low-quality green innovation. In exploring the underlying mechanisms, Liu et al. (2023) find that digital transformation promotes green innovation by increasing innovation resource investment and reducing debt costs. Considering board characteristics, existing literature shows that board gender diversity, board size, and education level positively moderate the relationship between digital transformation and green innovation (Lin & Xie, 2024). From a human capital perspective, Cheng et al. (2023) examine the impact of digital transformation on labor input and find an inverted U-shaped relationship, with labor input significantly enhancing internal corporate entrepreneurship.

4.1.3.3 Digital transformation and firm value

In this section, we focus on analyzing the 443 papers related to "digital transformation" and firm value." Specifically, we categorize the literature into total factor productivity and market value.

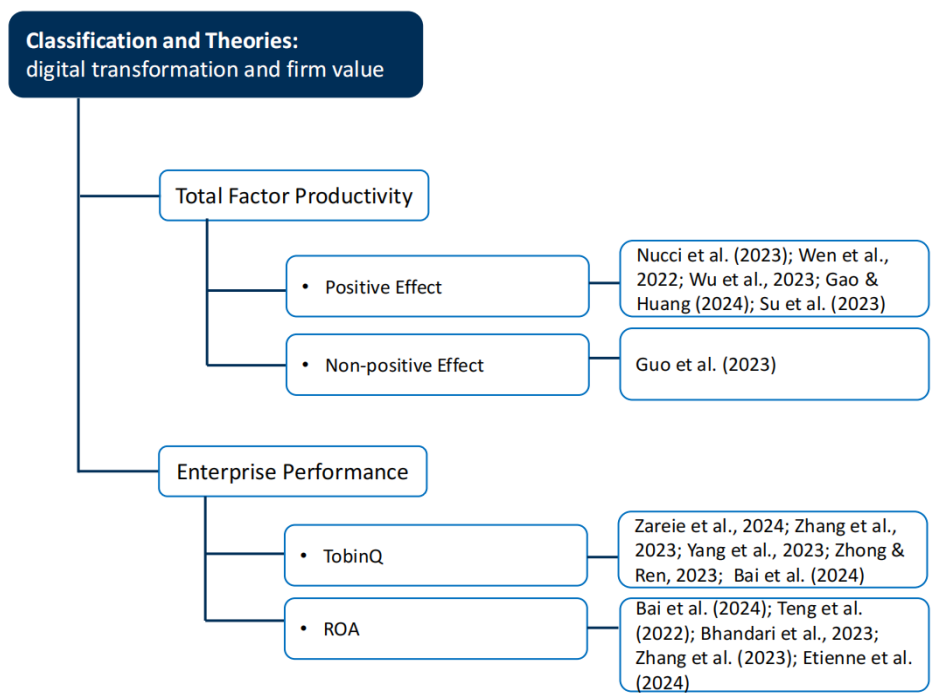

Fig. 12 Digital transformation and firm value.

**(1) Total factor productivity**

Nucci et al. (2023) find that digital transformation can improve production efficiency (Wen et al., 2022; Wu et al., 2023). However, Guo et al. (2023) find an inverted U-shaped relationship between digital transformation and total factor productivity, where both excessive and insufficient levels of digital transformation hinder production efficiency. Additionally, managerial myopia exacerbates this negative impact. Gao & Huang (2024) suggest that digital transformation improves green TFP mainly through improving resource allocation efficiency and stimulating green technology innovation. Su et al. (2023) find that the digital transformation of heavily polluting companies can effectively improve total factor productivity by increasing their level of green technology innovation and externally by increasing their willingness and capacity for corporate social responsibility. At the same time, digital transformation can improve total factor productivity by reducing cost stickiness, revealing the "black box" in which digital transformation affects the total factor productivity.

**(2) Enterprise performance**

From the perspective of enterprise risk management (ERM), Xu et al. (2024) divide enterprise risk management into strategic decision-making effectiveness and internal control based on the theory of resilient risk management. They find that digital transformation enhances firm value (Zareie et al., 2024; Zhang et al., 2023; Yang et al., 2023) by improving strategic effectiveness and internal control.

Distinguishing between short- and long-term firm value, Zhong & Ren (2023) find that digital transformation reduces short-term performance but enhances long-term value. Long et al. (2024) argue that digital transformation has significantly promoted the development of low-carbon cities, mainly through the promotion of green technology innovation, the expansion of the green investment scale, and the transformation of industrial structures. Feng & Nie (2024) find digital transformation significantly improves ESG performance, particularly for small-sized and non-state-owned firms. In addition, firms are also influenced by peer firms. The industry peer effect of digital transformation improves the environmental performance of companies by promoting innovation, while the regional peer effect improves their environmental performance by alleviating financing constraints (Ren et al., 2024). Bai et al. (2024) find that digital investment can impact intellectual capital through human and social capital (Guerra et al., 2023), enhancing enterprise value. Teng et al. (2022) find that the digital transformation of small and medium-sized enterprises positively correlates with business performance and has an inverted U-shaped relationship with innovation (Bhandari et al., 2023). Using ambidexterity theory, Zhang et al. (2023) classify digital transformation into exploitative and explorative types, both of which significantly enhance corporate performance, with business model innovation as a critical mediator.

Etienne et al. (2024) further show that the positive impact of SME digital transformation on performance depends heavily on firm characteristics, with rigid, radical-change-oriented SMEs seeing lower returns. Bai et al. (2024) find that digital transformation boosts investment in new technologies and R&D, improves the financing environment, reduces costs, and ultimately increases market value.

4.1.4 Digital transformation and others.

Zhang & Shen (2024) find that digital investment significantly increases the likelihood of hiring one of the Big Four accounting firms, indicating a growing demand for high-quality audit services. Deng et al. (2024) highlight the moderating role of digital transformation, showing that while executive turnover typically leads to a significant reduction in R&D investment, digital transformation mitigates this negative effect by improving communication efficiency and enhancing organizational learning, enabling firms to sustain R&D investment during executive transitions. Zhou et al. (2021) suggest that acquisitions related to digital technologies positively impact firms' innovation performance.

4.2 Analysis of research frontier

4.2.1 Timeline view of digital transformation

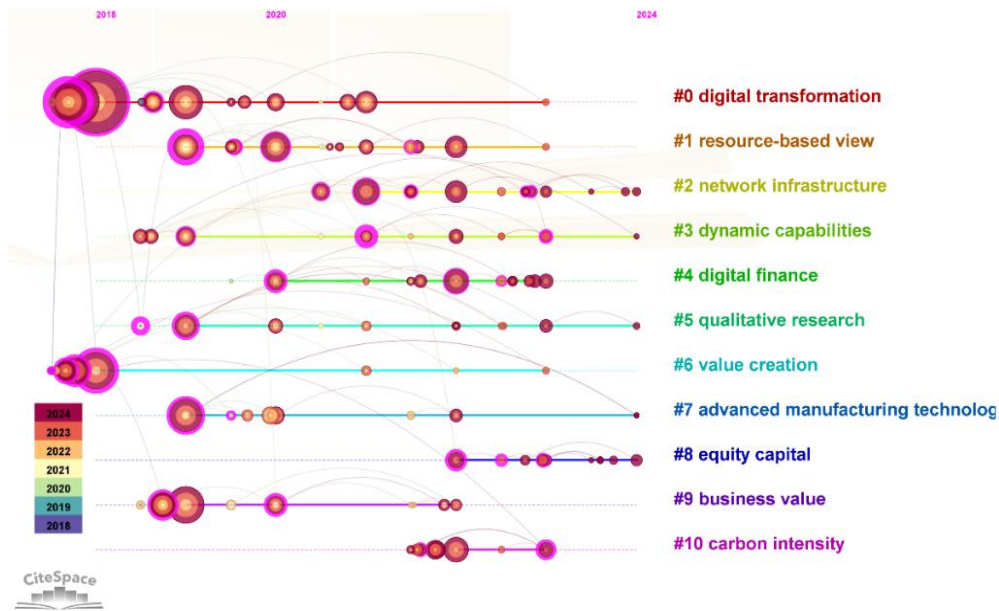

Fig. 13 Digital transformation literature keyword clustering time mapping.

Fig. 13 shows the keyword clustering timeline view of digital transformation research literature, which can further analyze the research hotspots and frontiers in this research area at different times. Cluster #0 (digital transformation) emerged rapidly in 2018, beginning widespread academic interest in this topic. Similarly, cluster #6 (value creation) also gained momentum around 2018. The purple halo around a node indicates the centrality of the keyword. We observe that cluster #3 (dynamic capabilities) had higher centrality from 2020 to 2024, indicating that this topic is a link between different research areas and contributed to the spread of interdisciplinary knowledge. Additionally, cluster #10 (carbon emission intensity) garnered significant academic attention around 2021 due to critical factors like climate risks and green credit. This hot topic continues to gain traction, with the integration of digitalization and green initiatives emerging as a potential research area.

4.2.2 Cluster analysis of co-citation of literature

The co-citation view of the literature reveals the clustering knowledge structure and evolving hot topics in digital transformation research. In Figure 14, each colored module border represents a thematic cluster within digital transformation research. The clusters in the graph are closely interconnected, evolving into 11 major sections. This finding aligns closely with the keyword clustering map described in Section 4.2.1, validating the efficiency of the clustering method. Additionally, it indicates that the hotspots and frontier directions regarding digital transformation are mainly consistent.

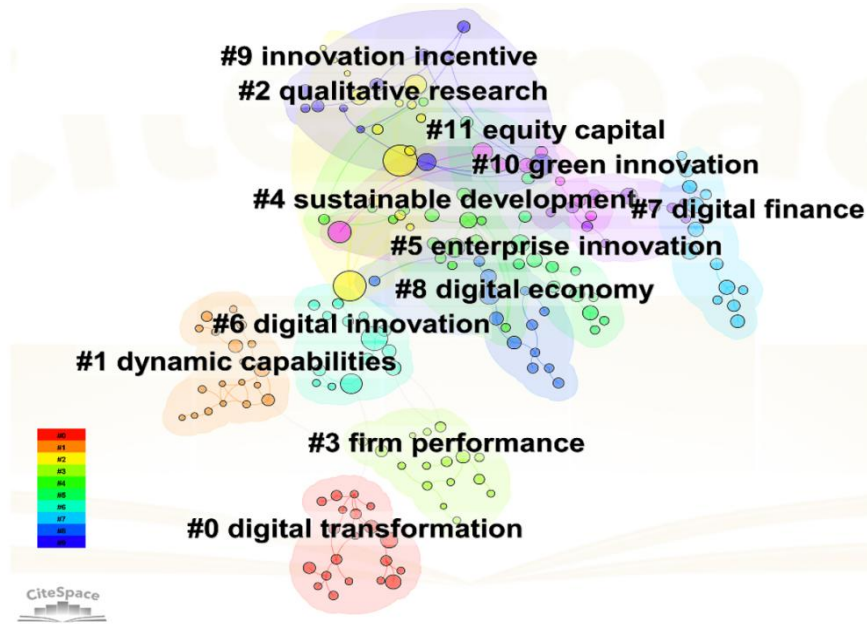

Fig. 14 Cluster analysis of co-citation of literature.

Table 4 shows the top 5 papers with the highest co-citation frequency as of the statistical date. Through a review of 282 works, Vial (2019) inductively constructs a framework of digital transformation comprising eight building blocks. Our framework positions digital transformation as a process in which digital technologies create disruptions that prompt organizations to respond strategically by altering their value creation paths while navigating structural changes and organizational barriers that influence the process's outcomes.

Digital transformation can drive business model innovation, fundamentally altering consumer expectations and behaviors. However, digital technologies also exert pressure on traditional firms. Verhoef et al. (2021) identify three stages of digital transformation: digitization, digitalization, and digital transformation, outlining the growth strategies for digital companies and the assets and capabilities required for successful transformation.

Warner & Wäger (2019) explored how traditional enterprises can establish the ability for dynamic digital transformation from a management perspective. This paper proposes a model consisting of nine micro-foundations to reveal the factors that promote and hinder the dynamic capabilities of digital transformation.

Matarazzo et al. (2021) show that, for SMEs, digital instruments contribute to innovation of their business model, creating new distribution channels and new ways to create and deliver value to customer segments. The results highlight the relevance of sensing and learning capabilities as triggers of digital transformation.

Wu et al. (2021) find that digital innovation significantly promotes corporate innovation performance, and the three subdimensions of dynamic capability, absorptive capacity, and innovation adaptability all play a mediating role in this relationship.

Table 4 Top 5 co-citation of literature.

| Freq | Title | Author | Year | Source |
|---|---|---|---|---|
| 219 | Understanding digital transformation: A review and a research agenda. | Vial G | 2019 | J STRATEGIC INF SYST |
| 181 | Digital transformation; A multidisciplinary reflection and research agenda. | Verhoef PC | 2021 | J BUS RES |
| 113 | Building dynamic capabilities for digital transformation: An ongoing process of strategic renewal | Warner KSR | 2019 | LONG RANGE PLANN |
| 85 | Digital transformation and customer value creation in Made in Italy SMEs: A dynamic capabilities perspective | Matarazzo M | 2021 | J BUS RES |
| 85 | Enterprise Digital Transformation and Capital Market Performance: Empirical Evidence from Stock Liquidity. | Wu F | 2021 | MANAGEMENT WORLD |

4.3 Analysis of bursts trend

This section generates a keyword bursts view for digital transformation research and extracts the top ten key literature with bursts intensity ranking, as shown in Figure 15. Nambisan (2017) believes there is a critical need for novel theorizing on digital innovation management. The authors incorporate four new theoretical logics into the framework of digital innovation management: (1) dynamic issues—solution design pairing; (2) social cognitive perception; (3) technological burdens and constraints; and (4) business processes.

Svahn (2017) presents a longitudinal case study of Volvo Cars' Connected car initiative. Combining extant literature with insights from the case, he argues that incumbent firms face four competing concerns—capability (existing versus requisite), focus (product versus process), collaboration (internal versus external), and governance (control versus flexibility)—and that these concerns are systemically interrelated. Firms must manage these concerns cohesively by continuously balancing new opportunities and established practices. To help managers respond to the opportunities and risk challenges of digital transformation, Hess (2016) answers the question of how German media companies can successfully achieve digital transformation.

Frank (2019), Dalenogare (2018), and Muller (2018) all explored digital technology based on Industry 4.0. Frank (2019) finds that Industry 4.0 is related to adopting front-end technologies, with intelligent manufacturing playing a central role. Dalenogare (2018) finds that Industry 4.0 technology can improve industrial efficiency in Brazil, while other emerging technologies have the opposite effect.

Müller (2018) finds Industry 4.0 encompasses three dimensions: high-grade digitization of processes, smart manufacturing, and inter-company connectivity. And this paper shows how Industry 4.0 affects the three business model elements of manufacturing SMEs – value creation, value capture, and value offer – by giving specific examples of business model innovation. Third, it demonstrates that a company's internal motivations and external pressures influence which business model elements are innovated. Fourth, the study delineates four SME categories to help managers evaluate their company's positioning towards Industry 4.0: craft manufacturers, preliminary stage planners, Industry 4.0 users, and full-scale adopters.

## Top 10 References with the Strongest Citation Bursts

| References | Year | Strength | Begin | End |
|---|---|---|---|---|
| Nambisan S, 2017, MIS QUART, V41, P223, DOI 10.25300/MISQ/2017/41:1.03, DOI | 2017 | 14.75 | 2019 | 2022 |
| Svahn F, 2017, MIS QUART, V41, P239, DOI 10.25300/MISQ/2017/41.1.12, DOI | 2017 | 10.56 | 2019 | 2022 |
| Hess T, 2016, MIS Q EXEC, V15, P123 | 2016 | 11.6 | 2020 | 2021 |
| Frank AG, 2019, INT J PROD ECON, V210, P15, DOI 10.1016/j.ijpe.2019.01.004, DOI | 2019 | 7.4 | 2020 | 2022 |
| Dalenogare LS, 2018, INT J PROD ECON, V204, P383, DOI 10.1016/j.ijpe.2018.08.019, DOI | 2018 | 7.04 | 2020 | 2022 |
| Müller JM, 2018, TECHNOL FORECAST SOC, V132, P2, DOI 10.1016/j.techfore.2017.12.019, DOI | 2018 | 6.9 | 2020 | 2022 |
| Foss NJ, 2017, J MANAGE, V43, P200, DOI 10.1177/0149206316675927, DOI | 2017 | 7.47 | 2021 | 2022 |
| Sebastian IM, 2017, MIS Q EXEC, V16, P197 | 2017 | 6.97 | 2021 | 2022 |
| Hinings B, 2018, INFORM ORGAN-UK, V28, P52, DOI 10.1016/j.infoandorg.2018.02.004, DOI | 2018 | 6.8 | 2021 | 2022 |
| Sklyar A, 2019, J BUS RES, V104, P450, DOI 10.1016/j.jbusres.2019.02.012, DOI | 2019 | 6.47 | 2021 | 2022 |

Fig. 15 Top 10 references with the strongest citation bursts.

# 5. Conclusion and discussion

5.1 Conclusion

We utilize two research methods, bibliometric and content analysis, to study the digital transformation research-related literature in the core collection of the WOS database and combine with CiteSpace software to qualitatively and quantitatively analyze the digital transformation research literature data from 2011 to 2024.

The topic of digital transformation emerged in 2015, with a significant literature explosion occurring from 2020 onwards. The number of published works has rapidly increased yearly, reaching 505 papers in 2023, making it a hot topic among scholars. China has made notable contributions to digital transformation, leading in publication volume with more articles than the United States (139) and the United Kingdom (112). However, some developed countries have established closer collaboration with international research than China. The primary journals publishing on digital transformation are predominantly high-impact international journals, with the Journal of Business Research leading the field as a core publication.

The research hotspots in digital transformation can be broadly categorized into three areas: (1) concepts and measurements of digital transformation (Cluster 3: Dynamic Capabilities, Cluster 5: Quantitative Measurement, Cluster 7: Advanced Technologies), (2) factors influencing corporate digital transformation (Cluster 2: Network Infrastructure Development, Cluster 4: Digital Finance, Cluster 8: Equity Capital), and (3) performance of digital transformation (Cluster 6: Value Creation, Cluster 9: Business Value, Cluster 10: Carbon Emission Intensity). This paper provides a detailed discussion of each sub-branches' theoretical foundations, theory, and future research directions.

5.2 Future research agenda

Based on a review of three WoS categories and ten sub-branches of digital transformation literature over the past 15 years, several insightful future research agendas are identified as follows:

- **Measurement and efficiency of digital transformation.** Digital transformation is a complex, systematic process. Although existing literature employs text analysis to measure, two issues persist. First, digital transformation encompasses the application of digital technologies and the integration with organization, management, and business. Relying on keywords fails to capture the full range of digital transformation actions undertaken by firms (Qi et al., 2021). Moreover, digital transformation methods differ across industries and ownerships, highlighting the need for industry-specific keyword dictionaries. Machine learning and natural language processing are valuable in analyzing unstructured data; thus, a key future research direction is leveraging large language models to create a more precise digital transformation framework.

- **Factors influencing digital transformation.** In the realm of the digital economy, developing an effective system for data asset protection represents a significant agenda for future research. Previous studies indicate that multinational enterprises (MNEs) have significantly altered their business strategies and structures to promote global integration (Garcia-Bernardo & Heemskerk, 2019). Consequently, the impact of infrastructure development on accelerating the digital transformation of multinational enterprises is an emerging research area. Moreover, in the study of informal institutions, culture is the core principle of corporate development and a basis for management decision-making. Examining how corporate culture affects digital transformation is another promising area for research. Furthermore, keyword cluster analysis indicates that "equity capital" has become increasingly significant since 2018, prompting an investigation into whether the role of venture capital in digital transformation mirrors its influence on innovation, which also merits investigation.

- **Digital transformation and financing.** There is limited literature on digital transformation and diversified financing. According to Modigliani and Miller (1984), capital structure is unrelated to corporate value. However, firms must establish an optimal capital financing structure due to frictions such as bankruptcy costs, agency costs, and information asymmetry. Digital resources, treated as a form of capital, influence this structure by being integrated into the production function. Future research directions may include examining the relationship between digital transformation and the concentration of corporate debt, the distinct effects of digital transformation on debt and equity financing decisions, and the interactive effects between digital technologies and capital structure. Additionally, "digital-green integration" has emerged as a significant research focus in recent years. Researchers could explore how digital transformation facilitates corporate green financing and the role of financial instruments, such as green bonds, in promoting corporate digitalization and sustainable development.

- **Digital transformation and investment.** While there is considerable literature on digital technology and corporate investment, research exploring the connection between digital transformation and labor investment efficiency within the framework of human capital is limited. Labor is a crucial production factor, particularly for knowledge-intensive firms (Ee et al., 2022). Previous studies have assumed that all companies invest in labor to some extent, with effective investment contributing to competitive success. However, inefficient labor investment is common in practice. Maintaining optimal levels of labor investment can significantly enhance operational efficiency (Pinnuck & Lillis, 2007; Caggese et al., 2019). Therefore, implementing digital transformation and improving labor investment efficiency are two key factors for companies to achieve sustainable competitive advantages. The relationship between digital transformation and corporate labor investment efficiency needs further investigation.

- **Digital transformation and firm value.** A key area for future research is how companies can balance financial and green performance during digital transformation amid rising low-carbon initiatives and increasing extreme weather events. Green transformation in manufacturing firms is urgent and challenging because of the dilemma with domestic resources and environmental constraints (Xie & Han, 2022). Digital technologies, including artificial intelligence, big data, and cloud computing, drive production processes (Dalenogare et al., 2018; Frank et al., 2019). Thus, an emerging research area is how digital technologies enable companies, particularly in manufacturing, to achieve cleaner development. Furthermore, since digital transformation represents a significant organizational change and involves complex cost-benefit trade-offs, incorporating digital resources as capital into the production function to analyze their role in supporting sustainable development within a multidimensional economy presents a potential research agenda.

# Reference:

[1] Bai, F., Shang, M., Huang, Y., & Liu, D. (2024). Digital investment, intellectual capital and enterprise value: evidence from China. Journal of Intellectual Capital, 25(1), 210-232.

[2] Bai, Y., Lu, C., Dong, X., & Li, Y. (2024). Role of collaborative governance in unlocking private investment in sustainable projects. Humanities and Social Sciences Communications, 11(1), 1-9.

[3] Bai, Z., Ban, Y., & Hu, H. (2024). Banking competition and digital transformation. Finance Research Letters, 61, 105068.

[4] Bhandari, K. R., Zámborský, P., Ranta, M., & Salo, J. (2023). Digitalization, internationalization, and firm performance: A resource-orchestration perspective on new OLI advantages. International Business Review, 32(4), 102135.

[5] Ceipek, R., Hautz, J., De Massis, A., Matzler, K., & Ardito, L. (2021). Digital transformation through exploratory and exploitative internet of things innovations: The impact of family management and technological diversification. Journal of Product Innovation Management, 38(1), 142-165.

[6] Chang, K., Li, J., & Xiao, L. (2024). Hear all parties: Peer effect of digital transformation on long-term firm investment in China. Managerial and Decision Economics, 45(3), 1242-1258.

[7] Chen, A., Zhao, D., & He, C. (2024). Firm digital investments and capacity utilisation: A perspective of factor inputs. Technological Forecasting and Social Change, 200, 123182.

[8] Chen, X., Cao, Y., Cao, Q., Li, J., Ju, M., & Zhang, H. (2023). Corporate financialization and digital transformation: evidence from China. Applied Economics, 1-16.

[9] Chen, X., Yan, Y., & Qiu, J. (2024). Can enterprise digital transformation reduce the reliance on bank credit? Evidence from China. Economic Modelling, 132, 106632.

[10] Chen, X., Zhou, P., & Hu, D. (2023). Influences of the ongoing digital transformation of the Chinese Economy on innovation of sustainable green technologies. Science of The Total Environment, 875, 162708.

[11] Chen, Y., Sun, R., & Zhang, T. (2024). Enterprise digital transformation and maturity mismatch: evidence from China. Applied Economics Letters, 1-5.

[12] Chen, Z., Xiao, Y., & Jiang, K. (2023). The impact of tax reform on firms' digitalization in China. Technological Forecasting and Social Change, 187, 122196.

[13] Cheng, Y., Zhou, X., & Li, Y. (2023). The effect of digital transformation on intrapreneurship in real economy enterprises: a labor input perspective. Management Decision.

[14] Cheng, Z., & Masron, T. A. (2023). Economic policy uncertainty and corporate digital transformation: evidence from China. Applied Economics, 55(40), 4625-4641.

[15] Deng, X., Hou, Q., & Shen, J. (2024). Can digital transformation reduce R&D investment disruption during the top executive transition period?. Managerial and Decision Economics, 45(4), 1901-1926.

[16] Dery, K., Sebastian, I. M., & van der Meulen, N. (2017). The digital workplace is key to digital innovation. MIS Quarterly Executive, 16(2).

[17] Dong, X., Meng, S., Xu, L., & Xin, Y. (2024). Digital transformation and corporate green innovation forms: evidence from China. Journal of Environmental Planning and Management, 1-29.

[18] Du, J., Shen, Z., Song, M., & Zhang, L. (2023). Nexus between digital transformation and energy technology innovation: An empirical test of A-share listed enterprises. Energy Economics, 120,

106572.

[19] Du, Z. Y., & Wang, Q. (2023). Unveiling the time-lag effects of digital transformation on financing constraints in Chinese listed enterprises: a study of automation and network technology. Technology Analysis & Strategic Management, 1-15.

[20] Du, Z., & Wang, Q. (2024). The power of financial support in accelerating digital transformation and corporate innovation in China: evidence from banking and capital markets. Financial Innovation, 10(1), 76.

[21] Etienne Fabian, N., Dong, J. Q., Broekhuizen, T., & Verhoef, P. C. (2024). Business value of SME digitalisation: when does it pay off more?. European Journal of Information Systems, 33(3), 383-402.

[22] Fan, J., Xiao, D., Xun, M., & Wang, C. (2024). Informal environmental regulation and enterprises digital transformation: A study based on the perspective of public environmental concerns. Ecological Indicators, 163, 112142.

[23] Fan, L., Ou, J., Yang, G., & Yao, S. (2024). Digitalization and outward foreign direct investment of Chinese listed firms. Review of International Economics, 32(2), 604-634.

[24] Feliciano-Cestero, M. M., Ameen, N., Kotabe, M., Paul, J., & Signoret, M. (2023). Is digital transformation threatened? A systematic literature review of the factors influencing firms' digital transformation and internationalization. Journal of Business Research, 157, 113546.

[25] Feng, H., Wang, F., Song, G., & Liu, L. (2022). Digital transformation on enterprise green innovation: Effect and transmission mechanism. International journal of environmental research and public health, 19(17), 10614.

[26] Feng, T., Appolloni, A., & Chen, J. (2024). How does corporate digital transformation affect carbon productivity? Evidence from Chinese listed companies. Environment, Development and Sustainability, 1-21.

[27] Feng, Y., & Nie, C. (2024). Digital technology innovation and corporate environmental, social, and governance performance: Evidence from a sample of listed firms in China. Corporate Social Responsibility and Environmental Management.

[28] Gao, L., & Huang, R. (2024). Digital transformation and green total factor productivity in the semiconductor industry: The role of supply chain integration and economic policy uncertainty. International Journal of Production Economics, 274, 109313.

[29] Gerth, A. B., & Peppard, J. (2016). The dynamics of CIO derailment: How CIOs come undone and how to avoid it. Business Horizons, 59(1), 61-70.

[30] Guerra, J. M. M., Danvila-del-Valle, I., & Méndez-Suárez, M. (2023). The impact of digital transformation on talent management. Technological Forecasting and Social Change, 188, 122291.

[31] Guo, C., Ke, Y., & Zhang, J. (2023). Digital transformation along the supply chain. Pacific-Basin Finance Journal, 80, 102088.

[32] Guo, L., Xu, L., Wang, J., & Li, J. (2023). Digital transformation and financing constraints of SMEs: evidence from China. Asia-Pacific Journal of Accounting & Economics, 1-21.

[33] Guo, X., Li, M., Wang, Y., & Mardani, A. (2023). Does digital transformation improve the firm's performance? From the perspective of digitalization paradox and managerial myopia. Journal of Business Research, 163, 113868.

[34] Hinings, B., Gegenhuber, T., & Greenwood, R. (2018). Digital innovation and transformation: An institutional perspective. Information and organization, 28(1), 52-61.

[35] Hu, D., Peng, Y., Fang, T., & Chen, C. W. (2022). The effects of executives' overseas background

on enterprise digital transformation: Evidence from China. Chinese Management Studies, 17(5), 1053-1084.

[36] Hu, Y., Che, D., Wu, F., & Chang, X. (2023). Corporate maturity mismatch and enterprise digital transformation: Evidence from China. Finance Research Letters, 53, 103677.

[37] Hua, Z., & Yu, Y. (2023). Digital transformation and the impact of local tournament incentives: Evidence from publicly listed companies in China. Finance Research Letters, 57, 104204.

[38] Huang, Y., Bai, F., Shang, M., & Liang, B. (2023). Catalyst or stumbling block: do green finance policies affect digital transformation of heavily polluting enterprises?. Environmental Science and Pollution Research, 30(38), 89036-89048.

[39] Jiang, K., Zhou, M., & Chen, Z. (2024). Digitalization and firms' systematic risk in China. International Journal of Finance & Economics.

[40] Jiang, Y., Zheng, Y., Fan, W., & Wang, X. (2024). Peer digitalization and corporate investment decision. Finance Research Letters, 61, 104995.

[41] Jin, X., Lei, X., & Wu, W. (2023). Can digital investment improve corporate environmental performance?-Empirical evidence from China. Journal of Cleaner Production, 414, 137669.

[42] Jović, M., Tijan, E., Aksentijević, S., & Pucihar, A. (2024). Assessing the Digital Transformation in the Maritime Transport Sector: A Case Study of Croatia. Journal of marine science and engineering, 12(4), 634.

[43] Khurram, M. U., Abbassi, W., Chen, Y., & Chen, L. (2024). Outward foreign investment performance, digital transformation, and ESG performance: Evidence from China. Global Finance Journal, 60, 100963.

[44] Kong, D., Liu, B., & Zhu, L. (2023). Stem CEOs and firm digitalization. Finance Research Letters, 58, 104573.

[45] Li, C., Liu, J., Liu, Y., & Wang, X. (2023). Can digitalization empowerment improve the efficiency of corporate capital allocation?—Evidence from China. Economic Analysis and Policy, 80, 1794-1810.

[46] Li, J., Zhang, G., Ned, J. P., & Sui, L. (2023). How does digital finance affect green technology innovation in the polluting industry? Based on the serial two-mediator model of financing constraints and research and development (R&D) investments. Environmental Science and Pollution Research, 30(29), 74141-74152.

[47] Li, M., Jiang, A., & Ma, J. (2023). Digital transformation and income inequality within enterprises– Evidence from listed companies in China. Pacific-Basin Finance Journal, 81, 102133.

[48] Li, Y., Xu, R., & Zhao, W. (2024). Beyond support and plunder? Government procurement and firm digital transformation in China. Applied Economics Letters, 1-5.

[49] Liao, F., Hu, Y., Sun, Y., & Ye, S. (2023). Does digital empowerment affect corporate green investment efficiency?. Environment, Development and Sustainability, 1-27.

[50] Lin, B., & Xie, Y. (2024). Impacts of digital transformation on corporate green technology innovation: Do board characteristics play a role?. Corporate Social Responsibility and Environmental Management, 31(3), 1741-1755.

[51] Liu, C., Zhang, W., & Zhu, X. (2022). Does digital transformation promote enterprise development?: evidence from Chinese A-share listed enterprises. Journal of Organizational and End User Computing (JOEUC), 34(7), 1-18.

[52] Liu, S., Wu, Y., Yin, X., & Wu, B. (2023). Digital transformation and labour investment efficiency: Heterogeneity across the enterprise life cycle. Finance Research Letters, 58, 104537.

[53] Liu, X. and Chen, Y., 2024. Can sharing auditors with customers improve suppliers digital transformation?. Frontiers in Psychology, 15, p.1336653.

[54] Liu, X., Liu, F., & Ren, X. (2023). Firms' digitalization in manufacturing and the structure and direction of green innovation. Journal of Environmental Management, 335, 117525.

[55] Liu, Y., & Wei, H. (2024). Digital transformation and enterprise financial asset allocation. Applied Economics, 1-18.

[56] Liu, Y., Cheng, J., & Dai, J. (2024). Harnessing digital transformation for green innovation in energy transition: a study on R&D investments and spatial spillover in China. Economic Change and Restructuring, 57(1), 13.

[57] Liu, Z., Zhou, J., & Li, J. (2023). How do family firms respond strategically to the digital transformation trend: Disclosing symbolic cues or making substantive changes?. Journal of Business Research, 155, 113395.

[58] Long, Y., Liu, L., & Yang, B. (2024). The effects of enterprise digital transformation on low-carbon urban development: Empirical evidence from China. Technological Forecasting and Social Change, 201, 123259.

[59] Luo, W., Yu, Y., & Deng, M. (2024). The impact of enterprise digital transformation on risk-taking: Evidence from China. Research in International Business and Finance, 69, 102285.

[60] Luo, Y., Cui, H., Zhong, H., & Wei, C. (2023). Business environment and enterprise digital transformation. Finance Research Letters, 57, 104250.

[61] Matarazzo, M., Penco, L., Profumo, G., & Quaglia, R. (2021). Digital transformation and customer value creation in Made in Italy SMEs: A dynamic capabilities perspective. Journal of Business research, 123, 642-656.

[62] Nadkarni, S., & Prügl, R. (2021). Digital transformation: a review, synthesis and opportunities for future research. Management Review Quarterly, 71, 233-341.

[63] Nie, J., Zhang, J., & Chang, X. (2024). Does social capital matter to firm digital transformation? Evidence from China. Finance Research Letters, 105636.

[64] Niu, Y., Wang, S., Wen, W., & Li, S. (2023). Does digital transformation speed up dynamic capital structure adjustment? Evidence from China. Pacific-Basin Finance Journal, 79, 102016.

[65] Nucci, F., Puccioni, C., & Ricchi, O. (2023). Digital technologies and productivity: A firm-level investigation. Economic Modelling, 128, 106524.

[66] Omrani, N., Rejeb, N., Maalaoui, A., Dabić, M., & Kraus, S. (2022). Drivers of digital transformation in SMEs. IEEE transactions on engineering management.

[67] Parra-Sánchez, D. T., & Talero-Sarmiento, L. H. (2024). Digital transformation in small and medium enterprises: a scientometric analysis. Digital Transformation and Society, 3(3), 257-276.

[68] Peng, C., Yang, S., & Jiang, H. (2022). Does digitalization boost companies' outward foreign direct investment?. Frontiers in psychology, 13, 1006890.

[69] Peng, H., Bumailikaimu, S., & Feng, T. (2024). The power of market: Venture capital and enterprise digital transforming. The North American Journal of Economics and Finance, 102218.

[70] Peng, Z., Huang, Y., Liu, L., Xu, W., & Qian, X. (2024). How government digital attention alleviates enterprise financing constraints: An enterprise digitalization perspective. Finance Research Letters, 67, 105883.

[71] Piluso, N. (2024). Tobin's Q and shareholder value: Does "shareholder return" impede investment?. Review of Financial Economics.

[72] Ren, X., Zeng, G., & Sun, X. (2023). The peer effect of digital transformation and corporate

environmental performance: empirical evidence from listed companies in China. Economic Modelling, 128, 106515.

[73] Saesen, J., Schmidt, C. V. H., & Strese, S. (2024). The more, the better: The influence of overconfident CEOs on their firms' digital orientation. Journal of Business Research, 183, 114809.

[74] Singh, N., Vishnani, S., Khandelwal, V., Sahoo, S., & Kumar, S. (2024). A systematic review of paradoxes linked with digital transformation of business. Journal of Enterprise Information Management, 37(4), 1348-1373. 数字化的综述

[75] Sklyar, A., Kowalkowski, C., Tronvoll, B., & Sörhammar, D. (2019). Organizing for digital servitization: A service ecosystem perspective. Journal of Business Research, 104, 450-460.

[76] Su, J., Wei, Y., Wang, S., & Liu, Q. (2023). The impact of digital transformation on the total factor productivity of heavily polluting enterprises. Scientific Reports, 13(1), 6386.

[77] Tang, M., Liu, Y., Hu, F., & Wu, B. (2023). Effect of digital transformation on enterprises' green innovation: empirical evidence from listed companies in China. Energy Economics, 128, 107135.

[78] Teng, X., Wu, Z., & Yang, F. (2022). Impact of the Digital Transformation of Small-and Medium-Sized Listed Companies on Performance: Based on a Cost-Benefit Analysis Framework. Journal of Mathematics, 2022(1), 1504499.

[79] Verhoef et al.(2021)-Digital transformation; A multidisciplinary reflection and research agenda. Journal of business research, 122(889).

[80] Vial, G. (2019). Understanding digital transformation: A review and a research agenda. Journal of Strategic Information Systems, 13-66.

[81] Wang, C., Li, J., Wang, Q., & Li, W. (2024). Stuck in the comfort zone? The influence of customer concentration on digital innovation in manufacturing firms. Innovation, 1-21.

[82] Wang, G., Lamadrid, R. L., & Huang, Y. (2024). Digital transformation and enterprise outward foreign direct investment. Finance Research Letters, 65, 105593.

[83] Wang, S., Li, X., Li, Z., & Ye, Y. (2023). The effects of government support on enterprises' digital transformation: Evidence from China. Managerial and Decision Economics, 44(5), 2520-2539.

[84] Wang, S., Wen, W., Niu, Y., & Li, X. (2024). Digital transformation and corporate labor investment efficiency. Emerging Markets Review, 59, 101109.

[85] Wang, W., Xiao, D., Wang, J., & Wu, H. (2024). The cost of pollution in the digital era: Impediments of air pollution on enterprise digital transformation. Energy Economics, 134, 107575.

[86] Warner, K. S., & Wäger, M. (2019). Building dynamic capabilities for digital transformation: An ongoing process of strategic renewal. Long range planning, 52(3), 326-349.

[87] Wen, H., Wen, C., & Lee, C. C. (2022). Impact of digitalization and environmental regulation on total factor productivity. Information Economics and Policy, 61, 101007. Wen, H., Wen, C., & Lee, C. C. (2022). Impact of digitalization and environmental regulation on total factor productivity. Information Economics and Policy, 61, 101007.

[88] Wen, H., Zhong, Q., & Lee, C. C. (2022). Digitalization, competition strategy and corporate innovation: Evidence from Chinese manufacturing listed companies. International Review of Financial Analysis, 82, 102166.

[89] Wen, J., & Deng, Y. (2023). How does intellectual property protection contribute to the digital transformation of enterprises?. Finance Research Letters, 58, 104340.

[90] Wu, F., Hu, H., Lin, H., & Ren, X. (2021). Enterprise Digital Transformation and Capital Market Performance: Empirical Evidence from Stock Liquidity. Journal of Management World, 37, 130-144+10.

[91] Wu, H., & Wang, Y. (2024). Digital transformation and corporate risk taking: Evidence from China. Global Finance Journal, 62, 101012.

[92] Wu, Y., Li, H., Luo, R., & Yu, Y. (2023). How digital transformation helps enterprises achieve high-quality development? Empirical evidence from Chinese listed companies. European Journal of Innovation Management.

[93] Xiao, D., Xu, J., & Li, Q. (2024). The "Double-Edged Sword" effect of air quality information disclosure policy—Empirical evidence based on the digital transformation of Chinese listed companies. Energy Economics, 133, 107513.

[94] Xiao, Y., Zhang, B., & Wang, H. (2024). Research on the impact of environmental regulations on green technological innovation in China from the perspective of digital transformation: a threshold model approach. Environmental Research Communications, 6(3), 035001.

[95] Xie, K., & Huang, W. (2023). The Impact of Digital Transformation on Corporate Tax Avoidance: Evidence from China. Discrete Dynamics in Nature and Society, 2023(1), 8597326.

[96] Xin, B., Liu, Y., & Xie, L. (2024). Data capital investment strategy in competing supply chains. Annals of Operations Research, 336(3), 1707-1740.

[97] Xu, C., Sun, G., & Kong, T. (2024). The impact of digital transformation on enterprise green innovation. International Review of Economics & Finance, 90, 1-12.

[98] Xu, G., Li, G., Sun, P., & Peng, D. (2023). Inefficient investment and digital transformation: What is the role of financing constraints?. Finance Research Letters, 51, 103429.

[99] Xu, N., Lv, W., & Wang, J. (2024). The impact of digital transformation on firm performance: a perspective from enterprise risk management. Eurasian Business Review, 1-32.

[100] Xu, W., Peng, Z., Feng, Z., & Peng, C. (2024). The impact of government digital attention on enterprise digital transformation: empirical insights from China. Applied Economics Letters, 1-6.

[101] Xu, Z., Meng, Q., & Wang, S. (2024). Digital transformation and innovation activities: evidence from publicly-listed firms in China. The European Journal of Finance, 1-17.

[102] Yang, G., Li, H., Nie, Y., Yue, Z., & Wang, H. (2023). Digital transformation and firm performance: the role of factor allocation. Applied Economics, 1-18.

[103] Yang, Y., Chen, W., & Yu, Z. (2023). Local government debt and corporate digital transformation: Evidence from China. Finance Research Letters, 57, 104282.

[104] Yao, W., Ni, M., Qian, Y., Yang, S., & Cui, X. (2024). CFO narcissism and corporate digital transformation☆. Finance Research Letters, 64, 105422.

[105] Yu, J., Xu, Y., Zhou, J., & Chen, W. (2024). Digital transformation, total factor productivity, and firm innovation investment. Journal of Innovation & Knowledge, 9(2), 100487.

[106] Zareie, M., Attig, N., El Ghoul, S., & Fooladi, I. (2024). Firm digital transformation and corporate performance: The moderating effect of organizational capital. Finance Research Letters, 61, 105032.

[107] Zhai, H., Yang, F., Gao, F., Sindakis, S., & Showkat, G. (2023). Digital Transformation and Over-Investment: Exploring the Role of Rational Decision-Making and Resource Surplus in the Knowledge Economy. Journal of the Knowledge Economy, 1-32.

[108] Zhang, K., & Bu, C. (2024). Top managers with information technology backgrounds and digital transformation: Evidence from small and medium companies. Economic Modelling, 132, 106629.

[109] Zhang, K., & Shen, Z. (2024). Enterprise Digital Investment and Auditor Choice: An Empirical Study Based on Chinese Listed Enterprises. Emerging Markets Finance and Trade, 1-16.

[110] Zhang, R., Gao, W., Chen, S., Zhou, L., & Li, A. (2024). Dose digital transformation contribute to

improving financing efficiency? Evidence and implications for energy enterprises in China. Energy, 300, 131271.

[111]Zhang, S., Lai, Y., Li, H., Wang, S., & Ru, X. (2024). Does digital transformation drive enterprise green investment? Evidence from Chinese listed firms. Journal of Environmental Planning and Management, 1-34.

[112]Zhang, X., Li, J., Xiang, D., & Worthington, A. C. (2023). Digitalization, financial inclusion, and small and medium-sized enterprise financing: Evidence from China. Economic Modelling, 126, 106410.

[113]Zhang, Y., Li, R., & Xie, Q. (2023). Does digital transformation promote the volatility of firms' innovation investment?. Managerial and Decision Economics, 44(8), 4350-4362.

[114]Zhang, Y., Ma, X., Pang, J., Xing, H., & Wang, J. (2023). The impact of digital transformation of manufacturing on corporate performance—The mediating effect of business model innovation and the moderating effect of innovation capability. Research in International Business and Finance, 64, 101890.

[115]Zhang, Y., Meng, Z., & Song, Y. (2023). Digital transformation and metal enterprise value: Evidence from China. Resources Policy, 87, 104326.

[116]Zhang, Z., Lu, Y., & Wang, H. (2024). The impact of management power on digital transformation. Asia Pacific Journal of Management, 1-25.

[117]Zhang, Z., Su, Z., & Tong, F. (2023). Does digital transformation restrain corporate financialization? Evidence from China. Finance Research Letters, 56, 104152.

[118]Zhao, L., & Yuan, H. (2024). Digital transformation and the herd effect of corporate green investment. Finance Research Letters, 63, 105244.

[119]Zhao, S., Zhang, L., An, H., Peng, L., Zhou, H., & Hu, F. (2023). Has China's low-carbon strategy pushed forward the digital transformation of manufacturing enterprises? Evidence from the low-carbon city pilot policy. Environmental impact assessment review, 102, 107184.

[120]Zhao, S., Zhang, L., Peng, L., Zhou, H., & Hu, F. (2024). Enterprise pollution reduction through digital transformation? Evidence from Chinese manufacturing enterprises. Technology in Society, 77, 102520.

[121]Zhao, X., Sun, X., Zhao, L., & Xing, Y. (2022). Can the digital transformation of manufacturing enterprises promote enterprise innovation?. Business Process Management Journal, 28(4), 960-982.

[122]Zhao, Y., Ren, A., Lin, Z., & Miao, Z. (2024). Cultural Barriers to Digital Innovation: Understanding the Impact of Gambling Culture on Chinese Firms. Journal of the Knowledge Economy, 1-36.

[123]Zhong, X., & Ren, G. (2023). Independent and joint effects of CSR and CSI on the effectiveness of digital transformation for transition economy firms. Journal of Business Research, 156, 113478.

[124]Zhou, B., & Ge, J. (2023). Does corporate digitization affect investment efficiency? Evidence from China. Applied Economics Letters, 1-6.

[125]Zhou, J., Liu, C., Xing, X., & Li, J. (2021). How can digital technology-related acquisitions affect a firm's innovation performance?. International Journal of Technology Management, 87(2-4), 254-283.

[126]Zhu, C., Li, N., Ma, J., & Qi, X. (2024). CEOs' digital technology backgrounds and enterprise digital transformation: The mediating effect of R&D investment and corporate social responsibility. Corporate Social Responsibility and Environmental Management, 31(3), 2557-2573.

[127]Zhu, J., Baker, J. S., Song, Z., Yue, X. G., & Li, W. (2023). Government regulatory policies for

digital transformation in small and medium-sized manufacturing enterprises: an evolutionary game analysis. Humanities and Social Sciences Communications, 10(1), 1-18.

[128]Zhuo, C., & Chen, J. (2023). Can digital transformation overcome the enterprise innovation dilemma: Effect, mechanism and effective boundary. Technological Forecasting and Social Change, 190, 122378.

[129]Xu, G., Li, G., Sun, P., & Peng, D. (2023). Inefficient investment and digital transformation: What is the role of financing constraints?. Finance Research Letters, 51, 103429.

[130]Luo, S. (2022). Digital finance development and the digital transformation of enterprises: based on the perspective of financing constraint and innovation drive. Journal of Mathematics, 2022(1), 1607020.

[131]Li, S., Gao, L., Han, C., Gupta, B., Alhalabi, W., & Almakdi, S. (2023). Exploring the effect of digital transformation on Firms' innovation performance. Journal of Innovation & Knowledge, 8(1), 100317.

[132]Qi, I. D., Du, B., & Wen, W. (2021). Digital strategic change of state-owned enterprises: mission embeddedness and mode selection—a case study based on the digitalization typical practices of three central enterprises. Management World, 11, 137-158.

[133]Galdino, K. M., Rezende, S. F. L., & Lamont, B. T. (2019). Market and internationalization knowledge in entrepreneurial internationalization processes. International Journal of Entrepreneurial Behavior & Research, 25(7), 1580-1600.

[134]Myers, S. C., & Majluf, N. S. (1984). wHEN FIRMS HAVE INFORMATION THAT INVESTORS. Journal of Financial Economics, 13, 187-221.

[135]Pinnuck, M., & Lillis, A. M. (2007). Profits versus losses: Does reporting an accounting loss act as a heuristic trigger to exercise the abandonment option and divest employees?. The Accounting Review, 82(4), 1031-1053.

[136]Caggese, A., Cuñat, V., & Metzger, D. (2019). Firing the wrong workers: Financing constraints and labor misallocation. Journal of Financial Economics, 133(3), 589-607.

[137]Ee, M. S., Hasan, I., & Huang, H. (2022). Stock liquidity and corporate labor investment. Journal of Corporate Finance, 72, 102142.

[138]Xie, X., & Han, Y. (2022). How can local manufacturing enterprises achieve luxuriant transformation in green innovation? A multi-case study based on attention-based view. J. Manag. World, 38(3), 76-106.

[139]Dalenogare, L. S., Benitez, G. B., Ayala, N. F., & Frank, A. G. (2018). The expected contribution of Industry 4.0 technologies for industrial performance. International Journal of production economics, 204, 383-394.

[140]Frank, A. G., Dalenogare, L. S., & Ayala, N. F. (2019). Industry 4.0 technologies: Implementation patterns in manufacturing companies. International journal of production economics, 210, 15-26.

[141]Geng, Y., Xiang, X., Zhang, G., & Li, X. (2024). Digital transformation along the supply chain: Spillover effects from vertical partnerships. Journal of Business Research, 183, 114842.

[142]Nambisan, S., Lyytinen, K., Majchrzak, A., & Song, M. (2017). Digital innovation management. MIS quarterly, 41(1), 223-238.

[143]Müller, J. M., Buliga, O., & Voigt, K. I. (2018). Fortune favors the prepared: How SMEs approach business model innovations in Industry 4.0. Technological forecasting and social change, 132, 2-17.